\documentclass[a4paper,fleqn]{cas-dc}

\usepackage[numbers]{natbib}
\usepackage{subfigure}
\usepackage{amsmath}
\usepackage{cases}
\usepackage{threeparttable}
\usepackage{stfloats} 
\usepackage{bbding} 
\usepackage{framed}
\usepackage{pifont} 
\usepackage{amsfonts}
\usepackage{subeqnarray}
\usepackage{longtable}
\usepackage{supertabular}
\usepackage{setspace}
\usepackage{multirow}
\usepackage{float}
\usepackage[figuresright]{rotating}
\usepackage{booktabs}
\usepackage{mathrsfs}
\usepackage{booktabs}
\usepackage{enumitem}
\usepackage{graphicx}
\makeatletter
\newcommand{\nosemic}{\renewcommand{\@endalgocfline}{\relax}}
\newcommand{\dosemic}{\renewcommand{\@endalgocfline}{\algocf@endline}}
\let\oldnl\nl
\newcommand{\nonl}{\renewcommand{\nl}{\let\nl\oldnl}}
\makeatother
\usepackage[export]{adjustbox} 
\usepackage{booktabs}
\usepackage{setspace}
\usepackage{xcolor}
\usepackage{mathrsfs}
\usepackage{amsmath}
\usepackage{array}
\usepackage{amssymb}
\usepackage{amsthm}
\usepackage{microtype}
\usepackage{url}
\usepackage{amsfonts,amssymb}
\usepackage{dsfont}
\usepackage{mathtools}
\usepackage{geometry}
\def\tsc#1{\csdef{#1}{\textsc{\lowercase{#1}}\xspace}}
\tsc{WGM}
\tsc{QE}

\begin{document}


\let\WriteBookmarks\relax
\def\floatpagepagefraction{1}
\def\textpagefraction{.001}

\shorttitle{Data-driven Power Flow Linearization: Simulation}    

\shortauthors{M. Jia, G. Hug, N. Zhang, Z. Wang, Y. Wang, C. Kang}  

\title [mode = title]{Data-driven Power Flow Linearization: Simulation}  



%

\author[inst1]{Mengshuo Jia}[type=editor,      
      orcid=0000-0002-2027-5314]

\author[inst1]{Gabriela Hug}

\author[inst2]{Ning Zhang}

\author[inst3]{Zhaojian Wang}

\author[inst4]{Yi Wang}

\author[inst2]{Chongqing Kang}






\affiliation[inst1]{organization={Department of Information Technology and Electrical Engineering, ETH Zürich},
            addressline={Physikstrasse 3}, 
            postcode={8092}, 
            state={Zürich},
            country={Switzerland}}

\affiliation[inst2]{organization={Department of Electrical Engineering, Tsinghua University},
            addressline={Shuangqing Rd 30}, 
            postcode={100084}, 
            state={Beijing},
            country={China}}

\affiliation[inst3]{organization={Department of Automation, Shanghai Jiao Tong University},
            addressline={Dongchuan Rd 800}, 
            postcode={200240}, 
            state={Shanghai},
            country={China}}

\affiliation[inst4]{organization={Department of Electrical and Electronic Engineering, The University of Hong Kong},
            addressline={Pok Fu Lam}, 
            state={Hong Kong},
            country={China}}




\begin{abstract}
Building on the theoretical insights of Part I, this paper, as the second part of the tutorial, dives deeper into data-driven power flow linearization (DPFL), focusing on comprehensive numerical testing. The necessity of these simulations stems from the theoretical analysis's inherent limitations, particularly the challenge of identifying the differences in real-world performance among DPFL methods with overlapping theoretical capabilities and/or limitations. The absence of a comprehensive numerical comparison of DPFL approaches in the literature also motivates this paper, especially given the fact that over 95\% of existing DPFL studies have not provided any open-source codes. To bridge the gap, this paper first reviews existing DPFL experiments, examining the adopted test scenarios, load fluctuation settings, data sources,  considerations for data noise/outliers, and the comparison made so far. Subsequently, this paper evaluates a total of 44 methods, containing over 30 existing DPFL approaches, some innovative DPFL techniques, and several classic physics-driven power flow linearization methods for benchmarking. The evaluation spans various dimensions, including generalizability, applicability, accuracy, and computational efficiency, using numerous different test cases scaling from 9-bus to 1354-bus systems. The numerical analysis in this paper identifies and examines significant trends and consistent findings across all methods under various test cases. Meanwhile, it offers theoretical insights into phenomena like under-performance, failure, excessive computation times, etc. Overall, this paper identifies the differences in the performances of the wide range of DPFL methods, reveals gaps not evident from theoretical discussions, assists in method selection for real-world applications, and provides thorough discussions on open questions within DPFL research, indicating ten potential future directions. (Word Count: 9668).
\end{abstract}




\maketitle

\section{Introduction}\label{sec:Intro}

Linear power flow models are of critical importance in power systems computations, subject to extensive research and widespread application across academia and industry, unlocking markets worth trillions and impacting every global consumer \cite{9914682, molzahn2019survey, 23, 8_1}. The precision and computational efficiency of these linearization methods are pivotal for operating and planning power systems, particularly the systems with high penetrations of renewable energy due to the fast varying nature of the resulting power flows. Enhancing the accuracy and efficiency of linear power flow models is therefore not just a nice-to-have technical improvement but a significant advance towards a sustainable energy future.

Data-driven power flow linearization (DPFL) has emerged as a promising method for acquiring high-precision linear models under must relaxed conditions, e.g., no need to know the physical model of the power system. It is thus garnering wide attention \cite{Powertech}. Despite being in the developing stage, DPFL has already cultivated a substantial knowledge base. This two-part tutorial aims to provide a comprehensive examination of DPFL approaches.

The first part of this tutorial \cite{partI} offered a thorough classification and theoretical analysis of all existing DPFL methods, including their mathematical foundations, analytical solutions, and critical assessments of each method’s capabilities, limitations, and applicability. This work serves as a foundational guide, catering to both beginners and experts within this area, as well as professionals from other disciplines simply seeking reliable linearization techniques. 

Despite the thoroughness of the theoretical analysis in \cite{partI}, it has limitations: when many linearization methods have similar strengths and/or weaknesses, it is almost impossible to predict their differences in terms of practical performance. Hence, with only \cite{partI}, identifying the most suitable method for specific needs still remains difficult. More importantly, existing numerical comparisons in the literature do not fully show the whole picture regarding the actual performance of DPFL approaches. The lack of a clear understanding of the actual performance differences among existing DPFL methods could mask the problems that are not apparent from the theoretical analysis of the capabilities and limitations, obscure the judgment of researchers within the DPFL community, and complicate the selection of appropriate linearization methods for potential users from other research fields. 

Indeed, implementing a comprehensive comparison requires substantial efforts, owing to the lack of open-source codes for over 95\% of the related literature. Nevertheless, in order to clarify ambiguities, outline future research paths, and benefit the community, this paper, as the second part of the tutorial, intends to fill this gap. Specifically, this paper conducts exhaustive simulations for all DPFL methods, some newly introduced DPFL methods to showcase DPFL's modular nature, and several classical physics-driven power flow linearization (PPFL) approaches as benchmarks, totaling 44 methods. The major focus of this paper is a thorough assessment of these methods in terms of generalizability, applicability, accuracy, and computational efficiency. The evaluation outcomes also support the identification of potential future directions. The contributions of this paper are therefore threefold:
\begin{itemize}
  \item[(i)] A comprehensive review of existing DPFL experiments is presented, examining the adopted test scenarios, load fluctuation settings, data sources, and the considerations for data noise/outliers. The review also gives an overview over the existing comparisons made among DPFL approaches, outlines the capabilities and limitations of previous experiments, and demonstrates the critical need for a comprehensive numerical comparison of all DPFL approaches. 
  \item[(ii)] An exhaustive numerical simulation of 44 linearization methods is conducted, including 36 existing DPFL approaches, four newly developed DPFL methods, and four classic PPFL algorithms. A detailed comparative analysis of these 44 methods is presented, discussing their generalizability, applicability, accuracy, and computational efficiency, thereby clarifying the actual performance of all the evaluated approaches.
  \item[(iii)] An in-depth discussion regarding the open research questions is provided, outlining ten promising but challenging future directions for DPFL research, informed by the numerical findings gained here and the theoretical conclusions drawn from the first part of the tutorial \cite{partI}. 
\end{itemize}

The remainder of this paper is organized as follows: Section II introduces the 44 methods. Section III reviews existing experiments in DPFL. Section IV assesses the methods regarding their generalizability and applicability. Section V details the numerical evaluations in terms of accuracy and computational efficiency. Section VI discusses open questions in the fields of DPFL, summarizing possible future directions. Section VII concludes the paper. 

\vspace{0.5cm}
\noindent \textbf{Remark}: \textit{We have made every effort to accurately replicate the methods described in the original research papers. However, due to factors such as the absence of open-source code (with very few exceptions) and often incomplete details in the literature, we cannot assure that our implementations perfectly reflect the original authors' intentions, although when the details were particularly vague, we even have developed multiple versions of the methods, as shown in Table \ref{tab:Abbr} in the next section. Nevertheless, we acknowledge that it is impossible to create exact replicas of the methods as envisioned by their creators. Additionally, it is important to note that no method is without flaws. The analysis of limitations in this paper is not meant as criticism but as part of a thorough evaluation under certain cases with given hyperparameters.}

\section{Evaluated Methods}
Table \ref{tab:Abbr} enumerates the 44 evaluated methods, detailing for each the corresponding abbreviation, the training algorithm employed, and any supporting techniques utilized. The following points warrant attention.

Firstly, for the linearly constrained programming approaches, we also evaluate these methods without their key constraints, e.g., the bound constraints, coupling constraints, or structure constraints, in order to verify the added value of incorporating such constraints. It is crucial to recognize that, even without these key constraints, the resulting programming models remain different, attributed to the varied supporting techniques they incorporate.

Secondly, the first part of this tutorial \cite{partI} reveals the modular nature of DPFL studies, highlighting their flexibility in the assembly of various techniques to forge novel methodologies. In alignment with this paradigm, we introduce several methods previously unexplored within the DPFL domain, and include them into the following comparative analysis. These approaches include the least squares with pseudoinverse, least squares augmented by principal component analysis, generalized least squares with pseudoinverse, and a clustering-based version of the partial least squares\footnote{In the adaptation of the least squares method to incorporate pseudoinverse, the conventional inversion operation used in the ordinary least squares method is substituted with the Moore–Penrose inverse. This adjustment is designed to enhance the method's resilience to multicollinearity issues.

Similarly, for the generalized least squares method augmented with pseudoinverse, the initial iteration of the well-known feasible generalized least squares method is modified to employ the least squares with pseudoinverse instead of the ordinary least squares. This modification also aims to strengthen the method's ability to manage multicollinearity.

In the case of clustering-based partial least squares, the approach involves substituting the ordinary least squares component within the clustering-based least squares methodology (as discussed in Part I \cite{partI}) with the ordinary partial least squares. This change seeks to better accommodate the inherent nonlinear characteristics of AC power flows.

For details on the least squares with principal component analysis, the reader is referred to Appendix A.}. It is crucial to clarify that the objective of integrating these methods is not to argue their ``novelty/superiority'' over all pre-existing techniques. Rather, we intend to demonstrate the ease with which one can deviate from conventional paths to devise distinct methodologies. Notably, some of these introduced methods have demonstrated satisfying performance and rankings in subsequent evaluations. This outcome, particularly given the unsophisticated-designed nature of these approaches, suggests a high potential for further advancements in DPFL research.

Finally, our evaluation also encloses a selection of physics-driven power flow linearization (PPFL) methods, such as the classic DC model, the power transfer distribution factor model, the warm-start 1st-order Taylor approximation model (derived from the equations of nodal power injections in polar coordinates), and the decoupled linearized power flow model \cite{11_14}. Note that these PPFL methods are widely recognized and employed in both academic research and industry practices.

\begin{table*}[h]
\centering
  \scriptsize
  \renewcommand{\arraystretch}{1.64}
  \caption{Abbreviations for All the Evaluated DPFL and PPFL Approaches}
  \label{tab:Abbr}
  \setlength{\tabcolsep}{0.2mm}{
  \begin{threeparttable}
\begin{tabular}
{@{}m{1cm}<{\centering}m{1.5cm}<{\centering}m{10cm}<{\centering}m{5.3cm}<{\centering}@{}}
\toprule
\multicolumn{4}{c}{\textbf{Data-driven Power Flow Linearization Approaches}}                                                                                                                                                                                       \\ \hline
\textbf{Ref.}                      & \textbf{Abbreviation} & \textbf{Training Algorithm }                                                                                                                      & \textbf{Supporting Technique}                                                \\ \hline
\cite{18}                 & LS           & Ordinary Least Squares                                                                                                                   & Voltage Squaring                                                    \\
\cite{7, 13}              & LS\_SVD     & Least Squares with Singular Value Decomposition                                                                                          & -                                                                \\
\cite{7, 13}              & LS\_COD      & Least squares with Complete Orthogonal Decomposition                                                                                     & -                                                                \\
\cite{8,2}                & LS\_HBLD     & Least Squares with Huber Loss - Solved Directly                                                                                          & -                                                                \\
\cite{8,2}                & LS\_HBLE     & Least Squares with Huber Loss - Equivalent Convex   Transformation                                                                       & -                                                                \\
\cite{11}                 & LS\_TOL      & Total Least Squares                                                                                                                      & -                                                                \\
\cite{12}                 & LS\_CLS      & Clustering-based Least Squares                                                                                                           & -                                                                \\
\cite{1, korda2018linear} & LS\_LIFX     & Ordinary Least Squares                                                                                                                   & Dimension Lifting: Lift $\boldsymbol{x}$                           \\
\cite{1}                  & LS\_LIFXi    & Ordinary Least Squares                                                                                                                   & Dimension Lifting: Lift the $i$-th dimension of   $\boldsymbol{x}$ \\
\cite{18}                 & LS\_WEI      & Ordinary Least Squares                                                                                                                   & Voltage Squaring; Forgetting Factor                                 \\
\cite{liu2023data}        & LS\_REC      & Recursive Least Squares                                                                                                                  & -                                                                \\
\cite{10}                 & PLS\_SIMRX   & Ordinary Partial Least Squares with SIMPLS                                                                                               & -                                                                \\
\cite{19}                 & PLS\_BDL     & Ordinary Partial Least Squares                                                                                                           & Bundle Variables; Replace Slack Bus's Injection                                \\
\cite{19}                 & PLS\_BDLY2   & Ordinary Partial Least Squares                                                                                                           & Bundle Variables               \\
\cite{17}                 & PLS\_REC     & Recursive Partial Least Squares                                                                                                          & -                                                                \\
\cite{17}                 & PLS\_RECW    & Recursive Partial Least Squares                                                                                                          & Forgetting Factor                                                   \\
\cite{6}                  & RR           & Ordinary Ridge Regression                                                                                                                & -                                                                \\
\cite{6}                  & RR\_VCS      & Ordinary Ridge Regression                                                                                                                & Voltage-angle Coupling; Voltage Squaring                            \\
\cite{4}                  & RR\_KPC      & Clustering-based Ridge Regression                                                                                                        & Voltage Squaring                                                    \\
\cite{24}                 & RR\_WEI      & Locally Weighted Ridge Regression                                                                                                        & -                                                                \\
\cite{14}                 & SVR          & Ordinary Support Vector Regression                                                                                                       & Voltage   Squaring                                                  \\
\cite{33}                 & SVR\_CCP     & Chance-constrained Programming                                                                                                           & Grid Topology Integration                                           \\
\cite{27,28}              & SVR\_POL     & Support Vector Regression with Kernels                                                                                                   & -                                                                \\
\cite{34}                 & SVR\_RR      & Support Vector Regression with Regularization                                                                                            & -                                                                \\
\cite{9}                  & LCP\_BOX     & Linearly Constrained Program with Bound Constraints                                                                                    & Voltage Squaring                                                    \\
\cite{9}                  & LCP\_COU     & Linearly Constrained Program with Coupling Constraints                                                                               & Grid Topology Integration                                           \\
\cite{11}                 & LCP\_JGD     & Linearly Constrained Program with Structure Constraints                                                                              & Bundle Known and Unknown   Variables                                                                \\
\cite{3}                  & DRC\_XM      & Moment-based Distributionally Robust Chance-constrained   Program: Random $\boldsymbol{X}$                                & -                                                                \\
\cite{3}                  & DRC\_XYM     & Moment-based Distributionally Robust Chance-constrained Program: Random $\boldsymbol{X}$ and $\boldsymbol{Y}$     & -                                                                \\
\cite{3}                  & DRC\_XYD     & Divergence-based Distributionally Robust Chance-constrained Program: Random $\boldsymbol{X}$ and $\boldsymbol{Y}$  & -                                                                \\
\cite{16, 23}             & DC\_LS       & Ordinary Least Squares                                                                                                                   & Physical Model's Coefficient Optimization via LS                           \\
\cite{9960825}            & DLPF\_C      & Ordinary Least Squares                                                                                                                   & Physical Model's Error Correction via QR$^\dagger$                                  \\
\cite{wold1975path}                      & PLS\_NIP     & Ordinary Partial Least Squares with NIPALS                                                                                             & -                                                                \\
-                  & LCP\_BOXN     & Linearly Constrained Program without Bound Constraints                                                                                    & Voltage Squaring                                                    \\
-                  & LCP\_COUN     & Linearly Constrained Program without Coupling Constraints                                                                               & Grid Topology Integration                                           \\
-                & LCP\_JGDN     & Linearly Constrained Program without Structure Constraints                                                                              & Bundle Known and Unknown   Variables                                                               \\
-                      & LS\_PIN      & Least Squares with Pseudoinverse                                                                                                & -                                                                \\
-                      & LS\_PCA      & Least Squares with Principal Component Analysis                                                                                                             & -                                                                \\
-                 & LS\_GEN      & Generalized Least Squares with Pseudoinverse                                                                                                              & -                                                  \\
-                      & PLS\_CLS     & Clustering-based Partial Least   Squares                                                                                                 & -                                                                \\ 
\toprule
\multicolumn{4}{c}{\textbf{Physics-driven Power Flow Linearization Approaches}}                                                                                                                                                                                    \\ \hline
\textbf{Ref. }                     & \textbf{Abbreviation} & \multicolumn{2}{c}{\textbf{Approach}}                                                                                                                                                                                   \\ \hline
-                      & DC           & \multicolumn{2}{c}{Classic Direct Current Model}                                                                                                                                                                       \\
-                      & PTDF         & \multicolumn{2}{c}{Classic Power Transfer Distribution Factors}                                                                                                                                                        \\
-                      & TAY          & \multicolumn{2}{c}{Warm-start 1st order Taylor approximation}                                                                                                                                                  \\
\cite{11_14}              & DLPF         & \multicolumn{2}{c}{Decoupled Linearized Power Flow} \\ 
\bottomrule
\end{tabular}
\begin{tablenotes}
    \footnotesize
    \item[$_-$]: Refers to ``none'' hereafter. 
    \item[$\dagger$]: Refers to QR decomposition.
  \end{tablenotes}
  \end{threeparttable}}
\end{table*}

\begin{table*}[]
  \centering
  \scriptsize
  \renewcommand{\arraystretch}{1}
  \caption{Review of the Numerical Experiments in the Existing DPFL Literature}
  \label{tab:case_study}
  \setlength{\tabcolsep}{0.2mm}{
  \begin{threeparttable}
    \begin{tabular}{@{}m{1cm}<{\centering}m{2cm}<{\centering}m{3cm}<{\centering}m{3cm}<{\centering}m{1.4cm}<{\centering}m{1.5cm}<{\centering}m{1.2cm}<{\centering}m{1.6cm}<{\centering}m{2.5cm}<{\centering}ccclll@{}}
      \toprule
      \textbf{Ref.} & \textbf{Approach}                                                                                                                                                                                                                     & \textbf{Transmission Case}                                                                                                                           & \textbf{Distribution Case}                                                                                                                                                    & \textbf{Fluctuation}$^\diamond$ & \textbf{Data}$^\dagger$                                          & \textbf{ Noise}                             & \textbf{ Outlier}                            & \textbf{DPFL Comparison$^\ddagger$}    &  &  &  \\ \hline
      \cite{18}           & LS                                                                                                                                          & -                                                                                                                                            & Modified IEEE 123-bus   Single-phase Case \cite{18_23}                                                                                                                 & -                      & Real Measured Data from   \cite{18_24}                        & Gaussian Noise                                  & -                                             &-  &  &  &  \\\hline
       \cite{7,   13}      & LS\_COD                                                                                                                                                             & IEEE 5-,24-,30-,57-,118-,and   300-bus Cases; PEGASE 1354-bus Case; RTE 1888-bus Case; Polish 2383-bus Case                           & -                                                                                                                                                                     & 40\%--100\%                & Generated by MATLAB                                           & -                                            & -                                             & LS\_TOL \cite{11}, PLS\_BDL \cite{19}  &  &  &  \\\hline
       \cite{8,2}          & \makecell[c]{LS\_HBLD\\LS\_HBLE}                                                                                                                                                                                                       & 22-, 85-, and 141-bus Cases                                                                                                                   & IEEE 13-, 123-, 8500-bus   Three-phase Cases                                                                                                                           & 50\%--150\%                & Generated by MATLAB and OpenDSS                               & -                                            & Generated from  [0 p.u., 3 p.u.]      &  -    &  &  &  \\\hline
      \cite{11}           & LS\_TOL                                                                                                                & IEEE 9-, 14-, 33-, and 57-bus Cases; NREL 118-bus Case; Case of Henan   province, China & -                                                                                                                                                                     & 80\%-120\%                & Partially Generated by MATLAB;   Partially from \cite{1_33}   & Gaussian Noise                                  & -                                             &PLS\_BDL \cite{19}  &  &  &  \\\hline
      \cite{12}           & LS\_CLS                                                                                                                                                                                                                     & -                                                                                                                                            & IEEE 123-bus Single-phase Case                                                                                                                                                      & -                      & Generated by GridLAB-D                                        & -                                            & -                                             &-  &  &  &  \\\hline
      \cite{1, korda2018linear}            & \makecell[c]{LS\_LIFX\\LS\_LIFXi}                                                                                                                                                                                                      & IEEE 5, 24, 30, 57, and 118-bus Cases; NREL 118-bus Case \cite{1_33}                                                                          & IEEE 33-bus Single-phase Case; Modified IEEE 123-bus Single-phase Case \cite{1_32}                                                                                 & 50\%--400\%                & Generated by Undisclosed Software                             & Gaussian Noise                                  & -                                             & PLS\_BDL \cite{19} &  &  &  \\\hline
      \cite{liu2023data}          & LS\_REC                                                                                                                                        & IEEE 118-bus \cite{birchfield2016grid} and Polish 2383-bus Cases \cite{zimmerman2016matpower}                                                                                                                                   & -                                                                                          & 80\%--120\%                      & Generated by MATPOWER and OpenDSS                        & Gaussian Noise                                  & Claimed to Consider but Details Undisclosed                                              & LS \cite{18}, LS\_HBLD, LS\_HBLE \cite{8,2}, LS\_COD \cite{7, 13}  &  &  &  \\\hline
      \cite{10}           & PLS\_SIMRX                                                                                                                                                                & IEEE 30-, 57-, 118-, and 300-bus   Cases                                                                                                      & -                                                                                                                                                                     & -                      & Generated by Undisclosed Software                             & -                                            & -                                             & PLS\_BDL\cite{19}  &  &  &  \\\hline
      \cite{19}           & \makecell[c]{PLS\_BDL\\PLS\_BDLY2}                                                                                                            & IEEE 5-, 57-,  118-, and 300-bus Cases                                                                                                        & IEEE 33-bus Single-phase Case \cite{18_9};   Modified 123-bus Single-phase Case \cite{19_32}                                                                         & 80\%--120\%                & Generated by MATLAB                                           & Gaussian Noise                                  & -                                             & -  &  &  &  \\\hline
      \cite{17}           & \makecell[c]{PLS\_REC\\PLS\_RECW}                                                                                                                                                                  & -                                                                                                                                            & IEEE 33-bus Single-phase Case                                                                                                                                          & -                      & Generated by MATLAB                                           & -                                            & -                                             & -  &  &  &  \\\hline
      \cite{6}            & \makecell[c]{RR\\RR\_VCS}                                                                                                                                                                    & IEEE 14-, 30-, 47-, 118-, and   300-bus Cases                                                                                                 & IEEE 33- and 123-bus Single-phase   Cases                                                                                                                              & -                      & Generated by MATLAB                                           & -                                            & -                                             & PLS\_BDL \cite{19}, SVR\_POL\cite{27,28} &  &  &  \\\hline
      \cite{4}            & RR\_KPC                                                                                                                                                                 & -                                                                                                                                            & Modified IEEE 37-bus, \cite{wang2017linear}, IEEE 123-bus \cite{4_45}, and 606-bus \cite{4_41} Three-phase  Cases & -                      & Generated by MATLAB                                           & Gaussian Noise                                  & -                                             & PLS\_BDL \cite{19}  &  &  &  \\\hline
      \cite{24}           & RR\_WEI                                                                                                                                                                                                                      & 8-generator 36-bus Case of   PJM, the U.S.                                                                                                    & -                                                                                                                                                                     & -                      & Partially from PJM and Partially   from \cite{24_7}           & Gaussian Noise                                  & -                                             &-  &  &  &  \\\hline
            \cite{14}           & SVR                                                                                                                                                                                                 & -                                                                                                                                            & Modified IEEE 33-bus Three-phase   Case \cite{14}                                                                                                                      & 50\%--150\%                & Generated by MATLAB                                           & -                                            & Generated from  [0.8 p.u., 1.2 p.u.] & PLS\_BDL \cite{19}  &  &  &  \\\hline
          \cite{33}           & SVR\_CCP                                                                                                                                                                     & 6-bus Case \cite{33_9};   118-bus Case \cite{33_118}; Texas 2000-bus Case \cite{33_20}                                                    & -                                                                                                                                                                     & 70\%--130\%                & Generated by MATLAB                                           & -                                            & -                                             & LS$^\S$  &  &  &  \\\hline
    \cite{27,   28}     & SVR\_POL                                                                                                                                                       & -                                                                                                                                            & IEEE 123-bus Single-phase Case; Southern California Edison Single-phase  Case, the U S.                                                                                                       & -                      & Partially Generated by MATLAB;   Partially Real Measured Data & Noise from Undisclosed Model & Generated from Gaussian   Model             & LS$^\S$  &  &  &  \\\hline
    \cite{34}           & SVR\_RR                                                                                     & -                                                                                                                                            & IEEE 123-bus Three-phase   Case;  IEEE 33-bus Case                                                                                                                   & 50\%--150\%                & Generated by MATLAB                                           & -                                            & Generated from  [0 p.u., 1 p.u.]     & PLS\_BDL \cite{19}, RR$^\S$  &  &  &   \\ \hline
      \cite{9}            & \makecell[c]{LCP\_BOX\\LCP\_COU}                                                                                                                           & IEEE 118-bus Case                                                                                                                             & -                                                                                                                                                                     & -                      & Generated by MATLAB                                           & -                                            & -                                             & SVR$^\S$  &  &  &  \\\hline
      \cite{3}            & \makecell[c]{DRC\_XM\\DRC\_XYM\\DRC\_XYD}                                                                                                                                       & IEEE 118-bus Case; Polish   2383-bus Case                                                                                                   & IEEE 123-, 8500-bus Three-phase   Case                                                                                                                                 & 60\%--140\%                & Generated by MATLAB                                           & -                                            & -                                             & LS\_HBLD, LS\_HBLE \cite{8,2}, LS\_COD \cite{7, 13}    &  &  &  \\ \hline
            \cite{16,   23}     & DC\_LS                                                                                                                                                                                                                           & Tennessee Valley Authority (TVA)   Case of the U.S. \cite{16_36}                                                                              & -                                                                                                                                                                     & -                      & Real Measured Data from   \cite{16_36}                        & -                                            & -                                             &-  &  &  &  \\\hline
      \cite{9960825}            & DLPF\_C                                                                                                                                                                                           & IEEE 14-bus, New England 39-bus, European 89-bus, IEEE 118-bus, Illinois 200-bus, and European 1354-bus Cases \cite{7725528}                                                                                                                                 & -              & Public Data                                            & From U.S. National Renewable Energy Laboratory \cite{draxl2015overview}                                             & -                                             & - & -  &  &  \\
       \bottomrule
  \end{tabular}
  \begin{tablenotes}
    \footnotesize
    \item[$\diamond$]: Refers to the fluctuation range for nodal power injections within the training and testing dataset. 
  \item[$\ddagger$]: Refers to the numerical comparison with DPFL approaches \textbf{ONLY}. 
  \item[$\S$]: Refers to the evaluated methods that the corresponding reference was not explicitly given. 
  \end{tablenotes}
\end{threeparttable}}
  \end{table*}

\section{Review of Existing Experiments}

Before conducting evaluations of the methods listed in Table \ref{tab:Abbr}, we here aim to offer a detailed review of existing DPFL experiments in the literature, as depicted in Table \ref{tab:case_study}. This review intends to present the experimental accomplishments of previous DPFL studies, while simultaneously revealing the importance and need for an extensive numerical comparison of all DPFL methods. Below are the further discussions of Table \ref{tab:case_study}. 

Firstly, Table \ref{tab:case_study} indicates that both transmission and distribution grids were used as test cases to verify DPFL methods. While distribution grids differ from transmission grids in terms of symmetry and topology, DPFL methods are generally applicable to both types of systems, such as the methods in \cite{1,2,3,6,8}. The reasons are twofold. First, even if the three phases in a distribution systems are unbalanced, DPFL methods can still be implemented by either training one DPFL model for each phase \cite{29}, or training an overall DPFL model for all variables in three phases \cite{4, 8, 2, 14}. Note that the latter can generate a DPFL model reflecting the mutual influences between phases. Second, from a DPFL perspective, radial topologies do not pose any unique challenges compared to mesh topologies, as the difference is only in the number of dependent and independent variables. In summary, while distribution grids may have unbalanced characteristics and radial topologies, these attributes do not bring special difficulties to DPFL studies.


Secondly, Table \ref{tab:case_study} reveals that many evaluations solely depended on artificial data for training and testing, without considering the effects of noise and outliers on the data. This ideal testing environment is rarely found in real life, and the resulting conclusions may not hold in practice. To address this issue, it is recommended that synthetic data be injected with noise and outliers to mimic real-world scenarios.

Thirdly, as indicated in Table \ref{tab:case_study}, only a few studies report the load fluctuation range used in their simulations. This is worth mentioning because the accuracy of the DPFL model is highly dependent on the simulated fluctuations. E.g., a narrow fluctuation range typically leads to higher accuracy for the DPFL model. Without this information being made public, it is difficult to determine the reason for the high accuracy of the evaluated DPFL model.

Finally, Table \ref{tab:case_study} indicates which of the DPFL approaches have been evaluated in existing DPFL studies. These evaluations aimed to implement a comparative analysis between established and novel DPFL methods at that time. However, the scope of these comparisons is quite narrow, with only a few DPFL studies undertaking evaluations against a limited number of existing DPFL approaches (some works only conducted comparisons with PPFL methods). Such constrained comparisons fail to provide an in-depth understanding of the overall performance of DPFL methods. Consequently, a more exhaustive and inclusive comparison across all the DPFL approaches is clearly needed, in order to demonstrate their relative merits and limitations comprehensively.




\section{Generalizability and Applicability Evaluations}
Let us start with the generalizability and applicability evaluations of the various approaches. Table \ref{tab:Appli} gives an overview of the evaluation results, which are further discussed below. 

\noindent \textbf{Notation}: in a general sense, $V$ refers to the nodal voltage magnitude (``voltage'' hereafter), $\theta$ denotes the nodal voltage phase angle (``angle'' hereafter), $P/Q$ refers to the nodal active/reactive power injection, PF/QF corresponds to the active/reactive line flow from the ``from-bus'', and PT/QT denotes the active/reactive line flow from the ``to-bus''. More specifically, $V_i$ represents the voltage of node $i$, $\theta_{ref}$ denotes the angle of the slack bus, and $\theta_{ij}$ indicates the angle difference between nodes $i$ and $j$. Additionally, $R_{ij}=V_{i}V_{j}\cos{\theta_{ij}}$ and $C_{ij}=V_{i}V_{j}\sin{\theta_{ij}}$.

\subsection{Predictor and Response Generalizability}\label{sec:appli}
While PPFL methods are often constrained in their choice of predictors and responses due to specific physical formulations, DPFL methods generally offer a more flexible framework. This flexibility indicates the potential for DPFL approaches to accommodate arbitrary known variables (including $P$ and $Q$ of the PQ buses, $V$ of the slack and PV buses, and $\theta$ of the slack bus) as predictors, and arbitrary unknown variables (including $P$ of the slack bus, $Q$ of the slack and PV buses, $V$ of the PQ buses, $\theta$ of the PQ and PV buses, and all active/reactive line flows, i.e., PF, PT, QF, QT) as responses. However, due to various factors, not all DPFL methods can achieve this level of generalizability:

\begin{itemize}
    \item For the PLS\_BDL and PLS\_BDLY2 methods, using $V$, $P$, and $Q$ as predictors is enforced by the bundle strategy \cite{partI} they adopt. This strategy, designed to address variations in bus types, inherently constrains the choice of predictors by pre-defining known and unknown variables. 
    \item The RR\_VCS method employs the voltage-angle coupling technique to more effectively manage the inherent nonlinearity of AC power flows \cite{partI}. The physical coupling relationship of this technique requires using $V^2, P$, and $Q$ as predictors and $V^2, P$, and $Q$ as predictors while taking $V^2$, $R_{ij}$, and $C_{ij}$ for $\forall i, j$ as responses. 
    \item The LCP\_BOX and LCP\_JGD methods integrate physical knowledge of power flows by formulating constraints based on the Jacobian matrix derived from AC power flow equations expressed in polar coordinates \cite{partI}. Within this Jacobian matrix, $P$ and $Q$ are treated as known variables, while $V$ and $\theta$ are considered unknown variables. 
    \item The LCP\_COU method is specifically designed to linearly estimate the values of branch flows by leveraging the terminal voltages and angles as predictive variables \cite{partI}. Consequently, the method can only employ $V$ and $\theta$ as predictors, and treat PF, PT, QF, and QT as responses. 
    \item For the DC\_LS and DLPF\_C methods, since they incorporate the DC and DLPF models respectively into their framework \cite{partI}, they must align their selection of predictors and responses with the underlying physical models they adopt.
\end{itemize}

\begin{table*}[H]
\centering
  \scriptsize
  \renewcommand{\arraystretch}{1.8}
  \caption{Generalizability and Applicability Evaluations of DPFL Approaches}
  \label{tab:Appli}
  \setlength{\tabcolsep}{0.2mm}{
  \begin{threeparttable}
\begin{tabular} 
{@{}m{1.5cm}<{\centering}m{1.5cm}<{\centering}m{1cm}<{\centering}m{2.3cm}<{\centering}m{2.3cm}<{\centering}m{2cm}<{\centering}m{2cm}<{\centering}m{2.5cm}<{\centering}m{2cm}<{\centering}ccclll@{}}
\toprule

                              \textbf{Approach}   & \textbf{Goal}  & \textbf{Model Type} & \textbf{Predictor Generalizability}  & \textbf{Response Generalizability}                                                 & \textbf{Multicollinearity Applicability} & \textbf{Zero Predictor Applicability} & \textbf{Constant Predictor Applicability} & \textbf{Normalization Applicability} \\ \hline

LS                  & Build         & 1             & Arbitrary           & Arbitrary                                                         & {\color{red}\textbf{×}}                          & {\color{red}\textbf{×}}                         & \checkmark                                      & \checkmark                      \\ 
 LS\_SVD             & Build         & 1             & Arbitrary           & Arbitrary                                                         & {\color{red}\textbf{×}}                          & \checkmark                         & \checkmark                                      & \checkmark                      \\ 
 LS\_COD             & Build         & 1             & Arbitrary           & Arbitrary                                                         & \checkmark                          & \checkmark                         & \checkmark                                      & \checkmark                      \\ 
 LS\_HBLD            & Build         & 1             & Arbitrary           & Arbitrary                                                         & \checkmark                          & \checkmark                         & \checkmark                                      & \checkmark                      \\ 
 LS\_HBLE            & Build         & 1             & Arbitrary           & Arbitrary                                                         & \checkmark                          & \checkmark                        & \checkmark                                      & \checkmark                      \\ 
 LS\_TOL             & Build         & 1             & Arbitrary           & Arbitrary                                                         & {\color{red}\textbf{×}}                          & \checkmark                         & \checkmark                                      & \checkmark                      \\ 
LS\_CLS             & Build         & 1             & Arbitrary           & Arbitrary                                                         & {\color{red}\textbf{×}}                          & {\color{red}\textbf{×}}                         & \checkmark                                      & \checkmark                      \\ 
 LS\_LIFX            & Build         & 3             & Arbitrary           & Arbitrary                                                         & \checkmark                          & \checkmark                         & \checkmark                                      & \checkmark                      \\ 
 LS\_LIFXi           & Build         & 3             & Arbitrary           & Arbitrary                                                         & \checkmark                          & \checkmark                         & \checkmark                                      & \checkmark                      \\ 
 LS\_WEI             & Build         & 1             & Arbitrary           & Arbitrary                                                         & \checkmark                          & \checkmark                         & \checkmark                                      & \checkmark                      \\ 
 LS\_REC             & Update        & 1             & Arbitrary           & Arbitrary                                                         & {\color{red}\textbf{×}}                          & {\color{red}\textbf{×}}                         & \checkmark                                      & \checkmark                      \\  
 PLS\_SIMRX          & Build         & 1             & Arbitrary           & Arbitrary                                                         & \checkmark                          & \checkmark                         & \checkmark                                      & \checkmark                      \\ 
 PLS\_BDL            & Build         & 1             & $V, P, Q$           & Arbitrary                                                         & \checkmark                          & {\color{red}\textbf{×}}                         & {\color{red}\textbf{×}}                                      & \checkmark                      \\ 
 PLS\_BDLY2          & Build         & 1             & $V, P, Q$            & Arbitrary                                                         & \checkmark                          & {\color{red}\textbf{×}}                         & {\color{red}\textbf{×}}                                      & \checkmark                      \\ 
 PLS\_REC            & Update        & 1             & Arbitrary           & Arbitrary                                                         & \checkmark                          & \checkmark                         & \checkmark                                      & \checkmark                      \\ 
 PLS\_RECW           & Update        & 1             & Arbitrary           & Arbitrary                                                         & \checkmark                          & \checkmark                         & \checkmark                                      & \checkmark                      \\ 
 RR                  & Build         & 1             & Arbitrary           & Arbitrary                                                         & \checkmark                          & \checkmark                         & \checkmark                                      & \checkmark                      \\ 
 RR\_VCS             & Build         & 3             & $V^2, P, Q$         & $V^2$,   $R_{ij}$,  $C_{ij}$ & \checkmark                          & \checkmark                         & \checkmark                                      & {\color{red}\textbf{×}}                        \\ 
 RR\_KPC             & Build         & 2             & Arbitrary           & Arbitrary                                                         & \checkmark                          & \checkmark                         & \checkmark                                      & \checkmark                      \\ 
 RR\_WEI             & Build         & 1             & Arbitrary           & Arbitrary                                                         & \checkmark                          & \checkmark                         & \checkmark                                      & \checkmark                      \\ 
 SVR                 & Build         & 1             & Arbitrary           & Arbitrary                                                         & \checkmark                          & \checkmark                         & \checkmark                                      & \checkmark                      \\ 
 SVR\_CCP            & Build         & 1             & Arbitrary           & Arbitrary                                                         & \checkmark                          & \checkmark                         & \checkmark                                      & \checkmark                      \\ 
 SVR\_POL            & Build         & 3             & Arbitrary           & Arbitrary                                                         & \checkmark                          & \checkmark                         & \checkmark                                      & \checkmark                      \\ 
 SVR\_RR             & Build         & 1             & Arbitrary           & Arbitrary                                                         & \checkmark                          & \checkmark                         & \checkmark                                      & \checkmark                      \\ 
 LCP\_BOX            & Build         & 1             & $P,   Q$            & $V,   \theta$                                                     & \checkmark                          & \checkmark                         & \checkmark                                      & {\color{red}\textbf{×}}                      \\ 
LCP\_COU            & Build         & 1             & $V,   \theta$     & PF,   PT, QF, QT                                                  & \checkmark                          & \checkmark                         & \checkmark                                      & {\color{red}\textbf{×}}                      \\ 
LCP\_JGD            & Build         & 1             & $P,   Q$            & $V,   \theta$                                                     & \checkmark                          & \checkmark                         & \checkmark                                      & {\color{red}\textbf{×}}                      \\ 
 DRC\_XM             & Build         & 1             & Arbitrary           & Arbitrary                                                         & \checkmark                          & \checkmark                         & \checkmark                                      & \checkmark                      \\ 
 DRC\_XYM            & Build         & 1             & Arbitrary           & Arbitrary                                                         & \checkmark                          & \checkmark                         & \checkmark                                      & \checkmark                      \\ 
 DRC\_XYD            & Build         & 1             & Arbitrary           & Arbitrary                                                         & \checkmark                          & \checkmark                        & \checkmark                                      & \checkmark                      \\ 
 DC\_LS              & Build         & 1             & $P$       & $\theta$                                                               & \checkmark                          & \checkmark                         & \checkmark                                      & {\color{red}\textbf{×}}                      \\ 
 DLPF\_C             & Build         & 1             & $V,   \theta, P, Q$ & $V,   \theta$, PF, QF                                             & \checkmark                          & \checkmark                         & \checkmark                                      & {\color{red}\textbf{×}}  \\ 
 LS\_PIN             & Build         & 1             & Arbitrary           & Arbitrary                                                         & \checkmark                          & \checkmark                         & \checkmark                                      & \checkmark                      \\ 
 LS\_PCA           & Build         & 1             & Arbitrary           & Arbitrary                                                         & \checkmark                          & \checkmark                         & \checkmark                                      & \checkmark                      \\ 
 LS\_GEN             & Build         & 1             & Arbitrary           & Arbitrary                                                         & \checkmark                          & \checkmark                         & \checkmark                                      & \checkmark                      \\ 
PLS\_CLS             & Build         & 2             & Arbitrary           & Arbitrary                                                         & \checkmark                          & \checkmark                         & \checkmark                                      & \checkmark                      \\ 
 PLS\_NIP             & Build         & 1             & Arbitrary           & Arbitrary                                                         & \checkmark                          & \checkmark                         & \checkmark                                      & \checkmark                      \\ 
 DC & Build & 1 & $P$ & $\theta$ & \checkmark                          & \checkmark                         & \checkmark                                      & {\color{red}\textbf{×}}                      \\ 
PTDF & Build & 1 & $P$ & PF & \checkmark                          & \checkmark                         & \checkmark                                      & {\color{red}\textbf{×}}                      \\ 
TAY & Build & 1 & $P, Q$ & $V, \theta$ & \checkmark                          & \checkmark                         & \checkmark                                      & {\color{red}\textbf{×}}                      \\ 
DLPF & Build & 1 & $V,   \theta, P, Q$ & $V,   \theta$, PF, QF & \checkmark                          & \checkmark                         & \checkmark                                      & {\color{red}\textbf{×}}                      \\ 
\bottomrule
\end{tabular}
\begin{tablenotes}
    \footnotesize
    \item \textbf{Terminology}: The term ``build'' refers to the method designed to construct a linear model from a static, historical dataset of electrical measurements. Conversely, ``update'' denotes the strategy to refine an existing linear model by progressively incorporating new measurements. Furthermore, models 1, 2, and 3 align with the model classifications outlined in Part I of this tutorial; refer to the Problem Formulation part in Section 1 of \cite{partI} for more details.
  \end{tablenotes}
\end{threeparttable}}
\end{table*}

Note that the constrained flexibility in choosing predictors and responses results in notable limitations. First, these methods might not leverage all available known data for model training, leading to potential information loss. For instance, the DC\_LS method only uses measurements of $P$, disregarding a large amount of known voltage data. Second, the capability to predict unknown variables using the developed linear model may be restricted. For example, the LCP\_BOX method is restricted to calculating branch flow values, leading to a quite limited functional scope.

\subsection{Applicability to Cases with Multicollinearity}
The ordinary least squares method struggles with multicollinearity \cite{partI}. Methods LS, LS\_CLS, and LS\_REC share this limitation, since they are all based on the ordinary least squares framework. Additionally, LS\_SVD and LS\_TOL are also affected by this problem, as discussed in \cite{partI}. Subsequent experiments will numerically demonstrate their limitations in this context.

\subsection{Zero Predictor Applicability}
The issue of zero predictors arises when certain known variable measurements in the training dataset are consistently zero. A typical example is the inclusion of the slack bus angle in the predictor set, whose value is commonly set as zero and remains zero throughout. Other instances may involve PQ buses where active/reactive power consumption is zero during the measurement period. This situation leads to zero columns in the predictor dataset matrix (where columns represent different variables, and rows represent individual measurements). Not all DPFL methods can handle these zero columns effectively:
\begin{itemize}
    \item Methods based on ordinary least squares, including LS, LS\_CLS, and LS\_REC, have difficulties with zero predictors, as these zero columns render the Gram matrix of the predictor matrix non-invertible, thereby leading to the failure of these methods. 
    \item The methods PLS\_BDL and PLS\_BDLY2 also struggle with zero columns. This limitation stems from the bundle strategy they employ, as outlined in \cite{partI}. The presence of zero columns results in a crucial matrix (i.e., $\hat{\boldsymbol{\beta}}_{22}$ in \cite{partI}) becoming non-invertible, which in turn leads to the ineffectiveness of these methods in such cases, as the invertibility of that matrix is necessary. 
\end{itemize}

\begin{figure*}[h]
	\centering 
	\includegraphics[width=6.9in]{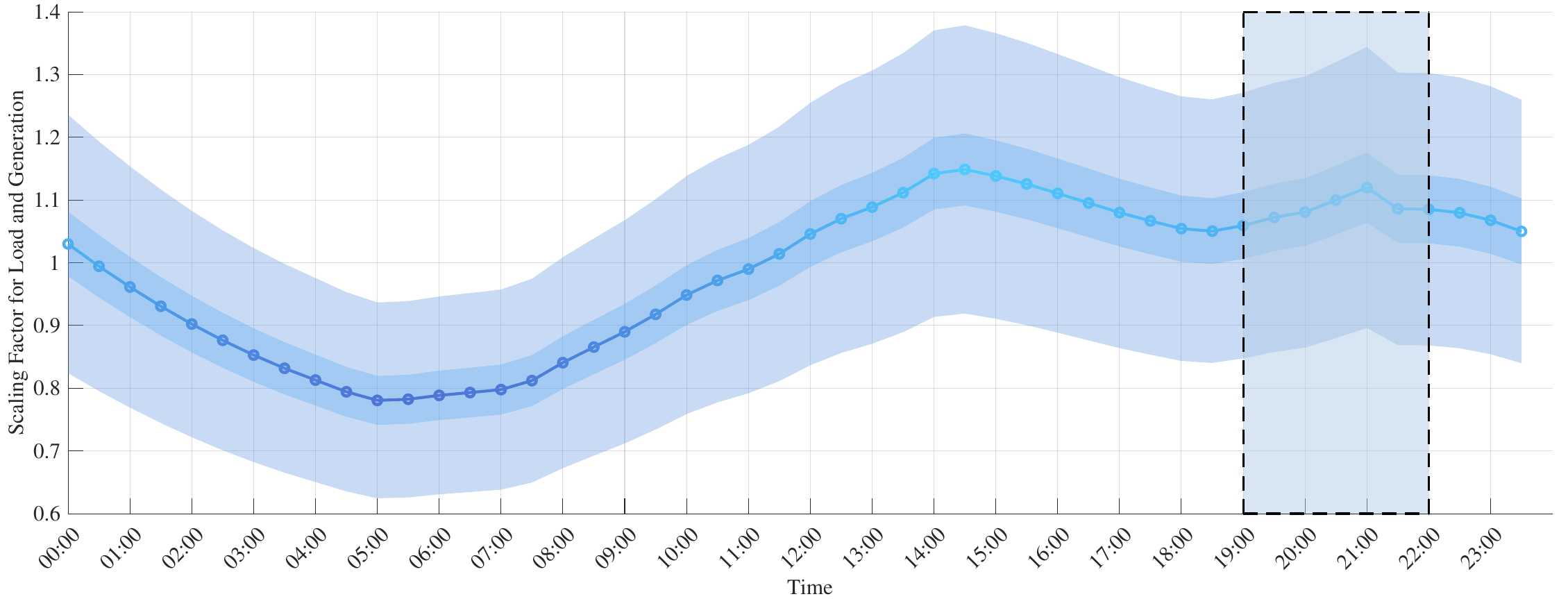} 
	\vspace{-0.4cm}
	\caption{Scaling factor curve for load and power generation variability. A specific focus is placed on the 19:00 to 22:00 time window, where significant changes occur. Two fluctuation intervals, represented by the two shaded zones, are depicted: a ``\textbf{Large Fluctuation Level}'' ranging from 80\% to 120\% and a ``\textbf{Small Fluctuation Level}'' from 95\% to 105\%. To alter loads/power generations, we first evenly divide the selected time window into $N_s$ time steps. At each time step, nodal load or power generation data is derived by multiplying the base value from the test system by the scaling factor plus a randomly sampled multiplier within the selected fluctuation ranges, e.g., within 80\% to 120\%. 
 }
	\label{Fig:LoadCurve} 
\end{figure*}

\subsection{Constant Predictor Applicability}
The constant predictor issue extends beyond the zero predictor problem, occurring when measurements of certain variables in the training dataset remain constant, not necessarily zero. A typical example is the fixed terminal voltages at PV buses, which normally stay constant over the measurement period, resulting in constant-value columns in the predictor matrix. 

Methods like PLS\_BDL and PLS\_BDLY2 cannot manage these constant columns due to, again, their reliance on the bundle strategy (detailed in \cite{partI}). Constant columns compromise the invertibility of a critical matrix in their strategy (again, $\hat{\boldsymbol{\beta}}_{22}$ in \cite{partI}), negatively affecting these methods' effectiveness. 

\subsection{Normalization Applicability}
As detailed in \cite{partI}, incorporating physical knowledge into DPFL methods can become problematic with datasets normalized via variance-scaling techniques, like unit-energy normalization, where each variable is normalized independently. This independent normalization disrupts the inherent physical relationships among variables, such as those represented in the Jacobian matrix or through coupling relationships, rendering methods like RR\_VCS, LCP\_BOX, LCP\_COU, LCP\_JGD, DC\_LS, and DLPF\_C inapplicable. 



\section{Numerical Evaluations}
For the numerical analysis, we chose a diverse range of transmission and distribution system models, spanning from 9-bus systems to 1354-bus systems, with 16 distinct configurations. In the following, we first detail the settings for each test case and the methods evaluated. Subsequently, we present the evaluation results with discussions from various perspectives. The open-source code for all the simulations in this paper can be found here \cite{code}. 

\subsection{Experiment Settings}

The detailed settings of all test cases are summarized in Table \ref{tab:cases}. To create the time-series variability in nodal loads and power generations in each test case, a typical daily load curve (in p.u.), shown in Fig. \ref{Fig:LoadCurve}, is used as a time-varying scaling factor to change both nodal demand and supply. We also introduce some randomness by adding different fluctuation levels to the scaling factor\footnote{Note that for a specific system we use the same scaling curve for all of the methods but across the different systems the randomness is different.}. Detailed descriptions are provided in the caption of Fig. \ref{Fig:LoadCurve}. The resulting supply and demand data are then inputted into AC power flow solvers to generate measurements like nodal voltage, angle, active and reactive power injections, and branch flows. All these measurements are further divided into training and testing datasets. The training dataset is used for DPFL methods to estimate the linear model, while the testing dataset is used to evaluate the linearization errors of DPFL and PPFL methods. The reader is referred to Table \ref{tab:para} for a thorough illustration of the settings for each method, including their predictors, responses, hyperparameters (or the cross-validation ranges\footnote{In this study, cross-validation is employed for hyperparameter tuning only in cases where the literature does not specify the hyperparameter values.}), datasets, solvers (when applicable), and additional relevant information. 

\begin{table*}[]
\centering
  \scriptsize
  \renewcommand{\arraystretch}{1.8}
  \caption{Settings for the Test Cases}
  \label{tab:cases}
  \setlength{\tabcolsep}{0.2mm}{
  \begin{threeparttable}
\begin{tabular} 
{@{}m{2.5cm}<{\centering}m{2.5cm}<{\centering}m{2.5cm}<{\centering}m{2.5cm}<{\centering}m{2.5cm}<{\centering}m{2.5cm}<{\centering}m{2cm}<{\centering}ccclll@{}}
\toprule
\textbf{Test Case} & \textbf{Training   Samples} & \textbf{Testing Samples} & \textbf{Fluctuation Level} & \textbf{Generator Terminal $V$} & \textbf{Noise Level$^\diamond$} & \textbf{Outlier Level$^\diamond$} \\ \hline
9-bus-S             & 150                         & 100                      & 95\%   - 105\%     & Fixed          & -                 & -                   \\
9-bus-L             & 150                         & 100                      & 80\%   - 120\%       & Fixed       & -                 & -                   \\
14-bus-S            & 200                         & 150                      & 95\%   - 105\%      & Fixed         & -                 & -                   \\
14-bus-L            & 200                         & 150                      & 80\%   - 120\%     & Fixed         & -                 & -                   \\
33-bus-L            & 300                         & 200                      & 80\%   - 120\%      & Fixed         & -                 & -                   \\
33-bus-L-NJ           & 300                         & 200                      & 80\%   - 120\%     & Fixed           & 45dB (Joint)$^\ddagger$     & -                   \\
33-bus-L-OJ           & 300                         & 200                      & 80\%   - 120\%      & Fixed          & -                 & 2\% (Joint)$^\dagger$      \\
33-bus-S            & 300                         & 200                      & 80\%   - 120\%       & Fixed        & -                 & -                   \\
33-bus-S-NI           & 300                         & 200                      & 80\%   - 120\%    & Fixed            & 45dB (Indivi)$^\P$     & -                   \\
33-bus-S-OI           & 300                         & 200                      & 80\%   - 120\%      & Fixed          & -                 & 2\% (Indivi)$^\S$      \\
39-bus-S            & 300                         & 200                      & 95\%   - 105\%       & Fixed        & -                 & -                   \\
39-bus-L           & 300                         & 200                      & 80\%   - 120\%       & Fixed         & -   & -                   \\
118-bus-S           & 400                         & 300                      & 95\%   - 105\%      & Fixed         & -                 & -                   \\
118-bus-L           & 400                         & 300                      & 80\%   - 120\%       & Fixed       & -                 & -                   \\
118-bus-L-NI          & 400                         & 300                      & 80\%   - 120\%     & Fixed         & 45dB (Indivi)  & -                   \\
118-bus-L-OI          & 400                         & 300                      & 80\%   - 120\%    & Fixed          & -                 & 2\% (Indivi)     \\
1354-bus-S          & 3000                        & 1000                     & 95\%   - 105\%       & Fixed        & -                 & -                   \\
1354-bus-L          & 3000                        & 1000                     & 80\%   - 120\%       & Fixed       & -                 & -       \\
\bottomrule
\end{tabular}
\begin{tablenotes}
    \footnotesize
    \item[$\diamond$]: Noise/outliers are only added to the training dataset; the testing data remains unpolluted; noise refers to the white Gaussian noise. 
    \item[$\ddagger$]: Within the training dataset's matrix, columns represent various variables, both known and unknown, while rows are indicative of distinct samples. The term ``Joint Noise'' refers to a scenario where each element within a row is subjected to an identical noise level of 45dB, a level suggested by \cite{11_11}. This implies that the entire system's data were measured by a single device at any given time instance, a premise that may not align with real-world practices. The inclusion of this scenario serves not only to illustrate that ``Joint Noise'' does not significantly impact the training performance (as shown later), but also to encourage researchers to clearly specify their methods of introducing noise into the dataset.
    \item[$\P$]: The term ``Indivi Noise'' means that each element within a row is affected by a unique instance of 45dB-level noise, implying that each variable within the system is measured by one device. This setting mirrors more realistic conditions and is anticipated to have a greater influence on the training performance.
  \item[$\dagger$]: Similarly, ``Joint Outlier'' involves randomly selecting 2\% of the rows within the training dataset and doubling their values simultaneously to simulate outliers, again under the unrealistic premise of single-device data collection. The inclusion of these scenarios aims to demonstrate their minimal impact on training outcomes, as evidenced in subsequent sections. It also highlights the importance of transparency in adding outliers to datasets. 
   \item[$\S$]: The term ``Indivi Outlier'' refers to the scenario where 2\% of elements within each column of the training dataset were randomly selected and doubled to create outliers, i.e., variables are measured by different devices. While this percentage seems small, when 2\% of elements in each column of a matrix are independently doubled to create outliers, the cumulative effect across multiple columns significantly amplifies the likelihood of a row being classified as an outlier, due to the increased chance of encountering at least one doubled element per row. 
  \end{tablenotes}
  \end{threeparttable} 
  }
\end{table*}

\begin{table*}[h]
\centering
  \scriptsize
  \renewcommand{\arraystretch}{1.63}
  \caption{Settings for the Evaluated Methods}
  \label{tab:para}
  \setlength{\tabcolsep}{0.2mm}{
  \begin{threeparttable}
\begin{tabular} 
{@{}m{1.5cm}<{\centering}m{3cm}<{\centering}m{1.5cm}<{\centering}m{10cm}<{\centering}m{2cm}<{\centering}ccclll@{}}
\toprule
\textbf{Approach} & \textbf{Predictor$^\diamond$}                & \textbf{Response$^\diamond$}     & \textbf{Miscellaneous}                                                                                & \textbf{Solver} \\ \hline
LS                & $P$, $Q$, $V^2$,   $\theta_{ref}$ & $V$, PF               & Data: Normalized$^\P$                                                                                                   & -            \\ 
LS\_SVD           & $P$, $Q$, $V^2$,   $\theta_{ref}$ & $V$, PF               & Data: Normalized                                                                                                   & -            \\ 
LS\_COD           & $P$, $Q$, $V^2$,   $\theta_{ref}$ & $V$, PF               & Data: Normalized                                                                                                   & -            \\ 
LS\_HBLD          & $P$, $Q$, $V^2$,   $\theta_{ref}$ & $V$, PF               & $\delta^{HUB} \in \{0.01, 0.02, \cdots, 0.05\}$; $N_{cv}=5$$^\dagger$; Data: Normalized                                          & FMINUNC         \\ 
LS\_HBLE          & $P$, $Q$, $V^2$,   $\theta_{ref}$ & $V$, PF               & $\delta^{HUB} \in \{0.01, 0.02, \cdots, 0.05\}$; $N_{cv}=5$ ; Data: Normalized                                           & MOSEK           \\ 
LS\_TOL           & $P$, $Q$, $V^2$,   $\theta_{ref}$ & $V$, PF               & Data: Normalized                                                                                                   & -            \\ 
LS\_CLS           & $P$, $Q$, $V^2$,   $\theta_{ref}$ & $V$, PF               & $k \in \{2, 3, \cdots, 10\}$; $N_{cv}=5$; Data: Normalized                                                               & -            \\ 
LS\_LIFX          & $P$, $Q$, $V^2$,   $\theta_{ref}$ & $V$, PF               & $f_{\text{lift}}$: Gauss; Data: Normalized                                                                               & -            \\ 
LS\_LIFXi         & $P$, $Q$, $V^2$,   $\theta_{ref}$ & $V$, PF               & $f_{\text{lift}}$: Gauss; Data: Normalized                                                                               & -            \\ 
LS\_WEI           & $P$, $Q$, $V^2$,   $\theta_{ref}$ & $V$, PF               & $\varpi$: 0.6 \cite{18}; Data: Normalized                                                                                & MOSEK           \\ 
LS\_REC           & $P$, $Q$, $V^2$,   $\theta_{ref}$ & $V$, PF               & New Data Share = 40\%$^\ddagger$; $\kappa = 0.99$; Data: Normalized                                                                & -            \\ 
PLS\_SIMRX        & $P$, $Q$, $V^2$,   $\theta_{ref}$ & $V$, PF               & $N_p =$ $\rm{Rank}(\boldsymbol{X})$; Data: Normalized                                                                       & -            \\ 
PLS\_BDL          & $P$, $Q$, $V^2$                   & $V$, PF              & $N_p =$ $\rm{Rank}(\boldsymbol{X})$; Data: Normalized                                                                        & -            \\ 
PLS\_BDLY2        & $P$, $Q$, $V^2$                   & $V$, PF               & $N_p =$ $\rm{Rank}(\boldsymbol{X})$; Data: Normalized                                                                        & -            \\ 
PLS\_REC          & $P$, $Q$, $V^2$,   $\theta_{ref}$ & $V$, PF               & New Data Share = 40\%; Data: Normalized                                                                                  & -            \\ 
PLS\_RECW         & $P$, $Q$, $V^2$,   $\theta_{ref}$ & $V$, PF               & New Data Share = 40\%; $\varpi = 0.6$ \cite{17}; Data: Normalized                                                       & -            \\ 
RR                & $P$, $Q$, $V^2$,   $\theta_{ref}$ & $V$, PF               & $\lambda=10^{-10}$; Data: Normalized                                                                                     & -            \\ 
RR\_VCS           & $P$, $Q$, $V^2$                   & $V^2, R_{ij}, C_{ij}$ & $\lambda=10^{-10}$; Data: Original$^\P$                                                                                     & -            \\ 
RR\_KPC           & $P$, $Q$, $V^2$,   $\theta_{ref}$ & $V$, PF               & $\lambda=10^{-10}$; $k \in \{2, 3, \cdots, 10\}$; $\eta \in   \{10^2, 10^3, \cdots, 10^5\}$; $N_{cv}=5$; Data: Normalized & -            \\ 
RR\_WEI           & $P$, $Q$, $V^2$,   $\theta_{ref}$ & $V$, PF               & $\lambda=10^{-10}$; $\tau \in \{0.1, 0.11, \cdots, 0.35\}$;   $N_{cv}=5$; Data: Normalized                               & -            \\ 
SVR               & $P$, $Q$, $V^2$,   $\theta_{ref}$ & $V$, PF               & $\epsilon=10^{-4}$; $\omega=10$; Data: Normalized                                                                       & GUROBI          \\ 
SVR\_CCP          & $P$, $Q$, $V^2$,   $\theta_{ref}$ & $V$, PF               & $\epsilon=10^{-4}$; $\omega=10$; $M=10^6$;  $\zeta^{CCP}_{j}=95\%$; Data: Normalized                                    & GUROBI          \\ 
SVR\_POL          & $P$, $Q$, $V^2$,   $\theta_{ref}$ & $V$, PF               & $\epsilon=10^{-4}$; Kernel Model: 3rd-order Polynomial; Data: Normalized                                                                 & FITRSVM         \\ 
SVR\_RR           & $P$, $Q$, $V^2$,   $\theta_{ref}$ & $V$, PF               & $\epsilon=10^{-4}$; $\omega=10$; $\lambda=10^{-4}$; Data: Normalized                                                     & GUROBI          \\ 
LCP\_BOX          & $P$, $Q$                          & $V$, $\theta$         & Bound: min/max coefficients of the 1st-order Taylor Approximation Model$^\S$; Data: Original                                                                                              & MOSEK           \\ 
LCP\_BOXN        & $P$, $Q$                          & $V$, $\theta$         & Data: Original                                                                                                   & MOSEK           \\ 
LCP\_COU          & $V$, $\theta$                     & PF        & $\delta^{LIN}=10^{-2}$; Data: Original                                                                               & MOSEK           \\ 
LCP\_COUN         & $V$, $\theta$                     & PF        & Data: Original                                                                                                   & MOSEK           \\ 
LCP\_JGD          & $P$, $Q$                          & $V$, $\theta$         & Structure: 1st-order Taylor Approximation Model; Data: Original                                                                                                   & SDPT3           \\ 
LCP\_JGDN          & $P$, $Q$                          & $V$, $\theta$         & Data: Original                                                                                                   & SDPT3           \\ 
DRC\_XM           & $P$, $Q$, $V^2$,   $\theta_{ref}$ & $V$, PF               & $\epsilon_j=10^{-4}$; $\zeta^{DRC}_{j}=95\%$; Data: Normalized                                                           & MOSEK           \\ 
DRC\_XYM          & $P$, $Q$, $V^2$,   $\theta_{ref}$ & $V$, PF               & $\epsilon_j=10^{-4}$; $\zeta^{DRC}_{j}=95\%$; Data: Normalized                                                           & MOSEK           \\ 
DRC\_XYD          & $P$, $Q$, $V^2$,   $\theta_{ref}$ & $V$, PF               & $\epsilon_j=10^{-4}$; $\zeta^{DRC}_{j}=95\%$; Data: Normalized                                                           & GUROBI          \\ 
DC\_LS            & $P$                               & $\theta$              & Data: Original                                                                                                   & -            \\ 
DLPF\_C           & $P$, $Q$, $V$,   $\theta_{ref}$ & $V$, PF               & Data: Original                                                                                                   & -            \\ 
LS\_PIN           & $P$, $Q$, $V^2$,   $\theta_{ref}$ & $V$, PF               & Data: Normalized                                                                                                   & -            \\ 
LS\_PCA           & $P$, $Q$, $V^2$,   $\theta_{ref}$ & $V$, PF               & $N_p \in \{40, 50, \cdots, 80\}$; $N_{cv}=5$; Data: Normalized                                                           & -            \\ 
LS\_GEN           & $P$, $Q$, $V^2$,   $\theta_{ref}$ & $V$, PF               & Covariance Estimation Model: ARIMA; Data: Normalized                                                                          & FGLS$^\&$            \\ 
PLS\_NIP          & $P$, $Q$, $V^2$,   $\theta_{ref}$ & $V$, PF               & Data: Normalized                                                                                                   & -            \\ 
PLS\_CLS          & $P$, $Q$, $V^2$,   $\theta_{ref}$ & $V$, PF               & $k \in \{2, 3, \cdots, 10\}$; $N_{cv}=5$; Data: Normalized                                                               & -            \\ 
DC                & $P$                               & $\theta$              & Data: Original                                                                                                  & -            \\ 
PTDF              & $P$                               & PF                    & Data: Original                                                                                                   & -            \\ 
TAY               & $P$, $Q$                          & $V$, $\theta$         & Data: Original                                                                                                   & -            \\ 
DLPF              & $P$, $Q$, $V$,   $\theta_{ref}$ & $V$, PF               & Data: Original                                                                                                   & -     \\ 
\bottomrule
\end{tabular}
\begin{tablenotes}
    \footnotesize
    \item[$\diamond$]: The default predictors and responses are \{$P$, $Q$, $V^2$, $\theta_{ref}$\} and \{$V$, PF\}, respectively, unless they do not apply to the approach (see Table \ref{tab:Appli}).  
    \item[$\P$]: The unit energy normalization is used for training and testing datasets by default, unless the approach is not applicable (see Table \ref{tab:Appli}).
  \item[$\dagger$]: $N_{cv}$ is the fold number for cross-validation, used for auto-tuning the parameter based on the given parameter set and the training dataset.
  \item[$\ddagger$]: ``New Data Share'' is the fraction of new data in the training set, used for recursive DPFL methods to update models in a point-wise manner. 
  \item[$\S$]: Every training data point has been used as a tangent point to find the min/max coefficients among all data points.
  \item[$_\&$]: $\rm{FGLS}$ in MATLAB has been modified by replacing the ordinary least squares with the least squares with Pseudoinverse, as previously mentioned.
  \end{tablenotes}
  \end{threeparttable} 
  }
\end{table*}

\subsection{Evaluation Overview}


We assess the methods using the previously described 16 test cases --- for each test case, the methods are ranked according to the active branch flow accuracy and separately according to the nodal voltage accuracy, resulting in 32 linearization rankings\footnote{Active branch flows and nodal voltages are two essential parameters frequently utilized in operation/planning/control models of power systems. The linearization accuracy evaluated by these two factors is thus crucial.}. To save space, we present here only four of these 32 rankings as representative examples. These rankings, illustrated in Figs. \ref{Fig:33-bus-S-PF} to \ref{Fig:39-bus-L-Vm}, cover various grid types, fluctuation levels, and data conditions, and will be discussed in subsequent sections. The other 28 evaulations are included in the supplementary material \cite{supplement}. Additionally, to concisely represent the comprehensive accuracy ranking information, we summarize all the accuracy results into Figs. \ref{Fig:rankPF} and \ref{Fig:rankVm}, respectively. The following accuracy analyses are largely drawn from these two figures. Lastly, Figure \ref{Fig:Efficiency} presents the computational efficiency for all evaluated methods.

\vspace{0.3cm}
\noindent \textbf{Remark}: \textit{The evaluation outcomes of various methods are influenced by method configurations (e.g., hyperparameters) and/or the settings of test cases (e.g., whether the dataset contains constant predictors such as the voltages of PV nodes). Given this, we intend not to argue that one method is universally superior/worse in accuracy or computational efficiency than others. Rather, our analysis below focuses on identifying and analyzing notable and consistent outcomes from numerous tests across various methods, in terms of their accuracy, inaccuracy, under-performance, failure, excessive computation times, etc. In the meantime, we dive into the underlying reasons to provide a theoretical explanation for these notable and consistent phenomena, instead of merely showing simulation outcomes. This strategy, complemented by the use of cross-validation for tuning parameters, may help to reduce the impact of method configurations and/or test case settings on the evaluation results. Nevertheless, it is important to acknowledge that such an impact can never be fully eliminated.}

\subsection{Failure Evaluation}
The figures below illustrate that certain evaluated methods encounter failures, which can be categorized into three distinct types:
\begin{itemize}
    \item \textbf{INA Type}: Certain methods cannot calculate specific responses due to limitations in response selection generalizability, leading to inapplicability (INA) failures. For example, the PTDF method cannot determine unknown voltages, leading to its failure when voltages serve as the evaluation benchmark.
    \item \textbf{NaN Type}: Some methods cannot produce a complete numerical coefficient matrix for the linear model, instead containing NaN values\footnote{NaN means ``not a number''. It occurs in case of undefined numeric results, such as $0 \div 0$, $0 \times \infty$, $\infty \div \infty$, etc.}, leading to NaN failures. This issue typically arises from attempts to invert a singular matrix during the training phase of a specific DPFL approach.
    \item \textbf{OOT Type}: Certain methods encounter computational limitations, leading to out-of-tolerance (OOT) failures. These limitations include exceeding the testing environment's available RAM (16 GB in our study), surpassing MATLAB's maximum array size (approximately 281.47 trillion elements in our study), or failing to complete training within a reasonable timeframe (set at one hour in our study). Such OOT failures underscore the significantly low computational efficiency of some approaches, especially optimization-based methods, when applied to larger test systems.  
\end{itemize}

In Figs. \ref{Fig:rankPF} and \ref{Fig:rankVm}, failures are explicitly identified and labeled. The following provides detailed explanations for these failures. 

\paragraph{INA-type Failure}~\\

First, For the calculation of active branch flows, methods LCP\_BOXN, LCP\_BOX, TAY (derived from the equations of nodal power injections in polar coordinates), LCP\_JGDN, and LCP\_JGD are not applicable, as discussed in Section \ref{sec:appli} and identified in Table \ref{tab:Appli}. 
    
Additionally, for the calculation of nodal voltages, methods PTDF,  LCP\_COUN, and LCP\_COU are not applicable, as explained in Section \ref{sec:appli} and emphasized in Table \ref{tab:Appli}. 

\vspace{1cm}
\paragraph{NaN-type Failure}~\\

First, the methods LS, LS\_CLS, and LS\_REC frequently face NaN-type failures due to their inability to address the singularity issue arising from multicollinearity. However, in some test cases where multicollinearity is less evident, these methods may not encounter such failures.
    
Second, the PLS\_BDLY2 method, which incorporates the active power injection at the slack bus into the predictor dataset, experiences a more severe multicollinearity issue compared to PLS\_BDL, as discussed in \cite{partI}. Consequently, PLS\_BDLY2 may face NaN-type failures in certain scenarios where PLS\_BDL does not. This also showcases the effectiveness of moving the active power injection at the slack bus to the response dataset.
    
Third, the failure of the PLS\_RECW method is primarily attributed to NIPALS algorithm used for component extraction and data decomposition for PLS\_RECW. Specifically, the issue arises when the predictor dataset's loading and weight matrices, produced by the NIPALS algorithm, exhibit linear dependency among their columns. This dependency results in a singular matrix upon their multiplication. Consequently, the linear model's coefficient matrix, derived from the inverse of this singular matrix, is rendered undefined, leading all coefficients to assume NaN values. A key underlying cause of this issue, especially when considering the successful performance of PLS\_REC and PLS\_NIP, is the implementation of the forgetting factor. In this study, a forgetting factor of 0.6, as referenced in \cite{17}, is applied. This factor effectively diminishes the influence of older samples due to its multiplication effect, as detailed in \cite{partI}. This mechanism is likely at the core of the observed numerical instability, as the failure issue of the PLS\_RECW method is resolved when the forgetting factor is adjusted to 0.9.

\paragraph{OOT-type Failure}~\\

First, methods DRC\_XYM and DRC\_XM, both moment-based, employ conic dual transformation to convert constraints into equivalent semi-definite constraints. Given the high computational demand of semi-definite programming, it is no surprise that these methods begin to exceed the RAM capacity of the testing device at a system size of 118 buses.

Second, methods LS\_LIFX and LS\_LIFXi utilize dimension lifting, significantly increasing the dataset size with the test case scale. At a system size of 1354 buses, the dimension-lifted dataset surpasses MATLAB's maximum array element limit.

Third, a range of methods, including LS\_HBLD, LS\_HBLE, LS\_WEI, SVR, SVR\_CCP, SVR\_RR, LCP\_BOX, LCP\_BOXN, LCP\_COU, LCP\_COUN, LCP\_JGD, LCP\_JGDN, and DRC\_XYD, either apply optimization-based approaches or use regression models but are solved via optimization programming. The number of decision variables (i.e., coefficient matrix elements of the linear model) exponentially increases with system size, rendering these methods unsolvable by commercial solvers at a system size of 1354 buses due to RAM constraints.

Lastly, methods SVR\_POL and LS\_GEN experience significant slowdowns at a system size of 1354 buses. The former is affected by its 3rd-order polynomial-kernel-based fitting process, while the latter's iterative process and the use of pseudoinverse result in substantial time consumption when dealing with large matrices.

\subsection{Accuracy Evaluation}

The following discussion on accuracy spans several areas, including a performance comparison between DPFL and PPFL methods, a general analysis of DPFL methods, and individual assessments of various DPFL approaches.

\subsubsection{DPFL Performance vs. PPFL Performance}
\begin{figure*}[H]
	\centering 
	\includegraphics[width=7.1in]{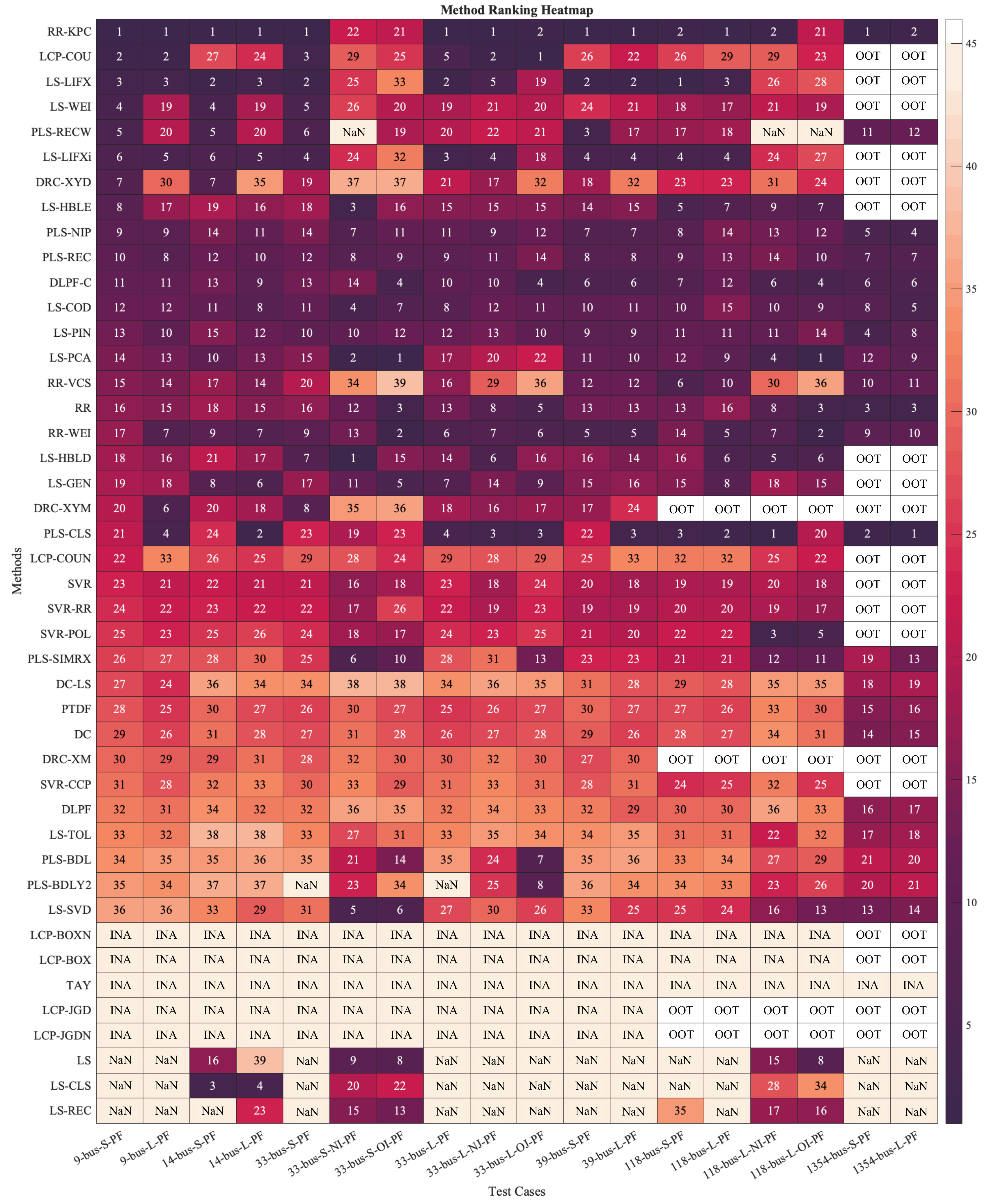} 
	\vspace{-0.4cm}
	\caption{Rankings of the linearization accuracy of 44 distinct methods for active branch flows across various test conditions. Rankings from 1 to 44 indicate each method's accuracy, with ``1'' being the most accurate. The visualization employs color intensity to signify rankings: darker shades represent superior rankings and enhanced accuracy. }
	\label{Fig:rankPF} 
\end{figure*}

Figs. \ref{Fig:rankPF} and \ref{Fig:rankVm} clearly demonstrate that DPFL approaches generally surpass the accuracy of the commonly used PPDL methods such as DC, PTDF, TAY, and DLPF. Notably, only TAY occasionally achieves higher rankings, such as third, fourth, or fifth in some cases, but other DPFL methods consistently outperform it. 

\begin{figure*}[H]
	\centering 
	\includegraphics[width=7in]{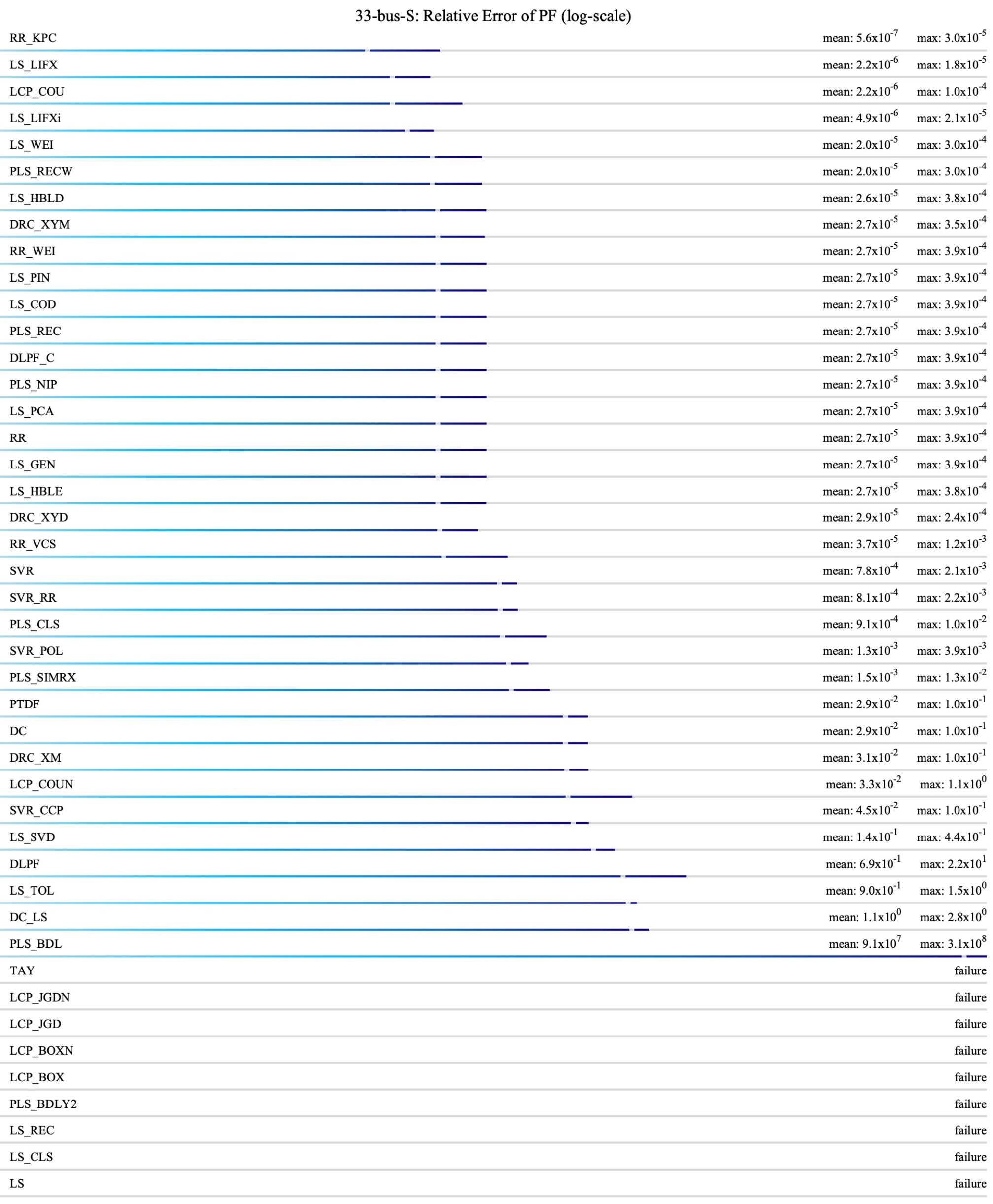} 
 \vspace{-0.3cm}
	\caption{Linearization accuracy ranking of active branch flows of all methods, evaluated under the 33-bus-S case. For each method, the relative error of active branch flow $i$ at time $t$ is computed by $\epsilon_{it} = |y_{it}^{true}-y_{it}^{method}|/|y_{it}^{true}|$. The mean relative error is thus the average of $\epsilon_{it}$ for $\forall i$ and $\forall t$, while the maximal relative error is the maximum of $\epsilon_{it}$ for $\forall i$ and $\forall t$. The methods are ranked based on their mean relative errors, which are depicted by the small white space within the error bars. The full length of each error bar indicates the maximum relative error observed. Note that these error bars are presented on a logarithmic scale. The computation of the mean/maximal relative error, as well as the graphical interpretations of the error bars and the small white spaces within them, are consistently applied across Figs. \ref{Fig:33-bus-S-NI-PF}, \ref{Fig:39-bus-S-Vm}, \ref{Fig:39-bus-L-Vm}, and the figures presented in the supplementary material \cite{supplement}}
	\label{Fig:33-bus-S-PF} 
\end{figure*}

\begin{figure*}[H]
	\centering 
	\includegraphics[width=7.1in]{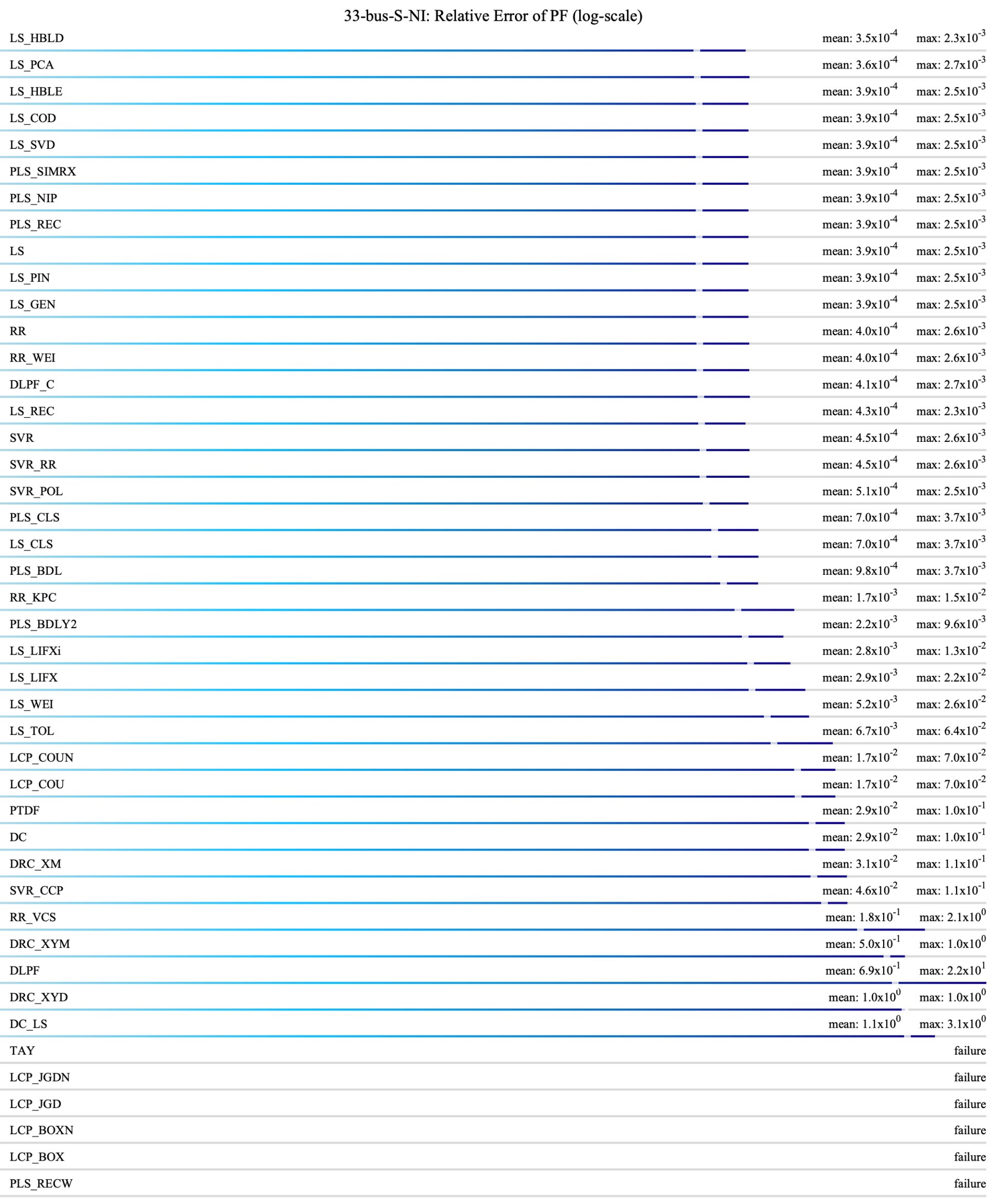} 
	\caption{Linearization accuracy ranking of active branch flows of all methods, evaluated under the 33-bus-S-NI case.}
	\label{Fig:33-bus-S-NI-PF} 
\end{figure*}

\begin{figure*}[H]
	\centering 
	\includegraphics[width=7.1in]{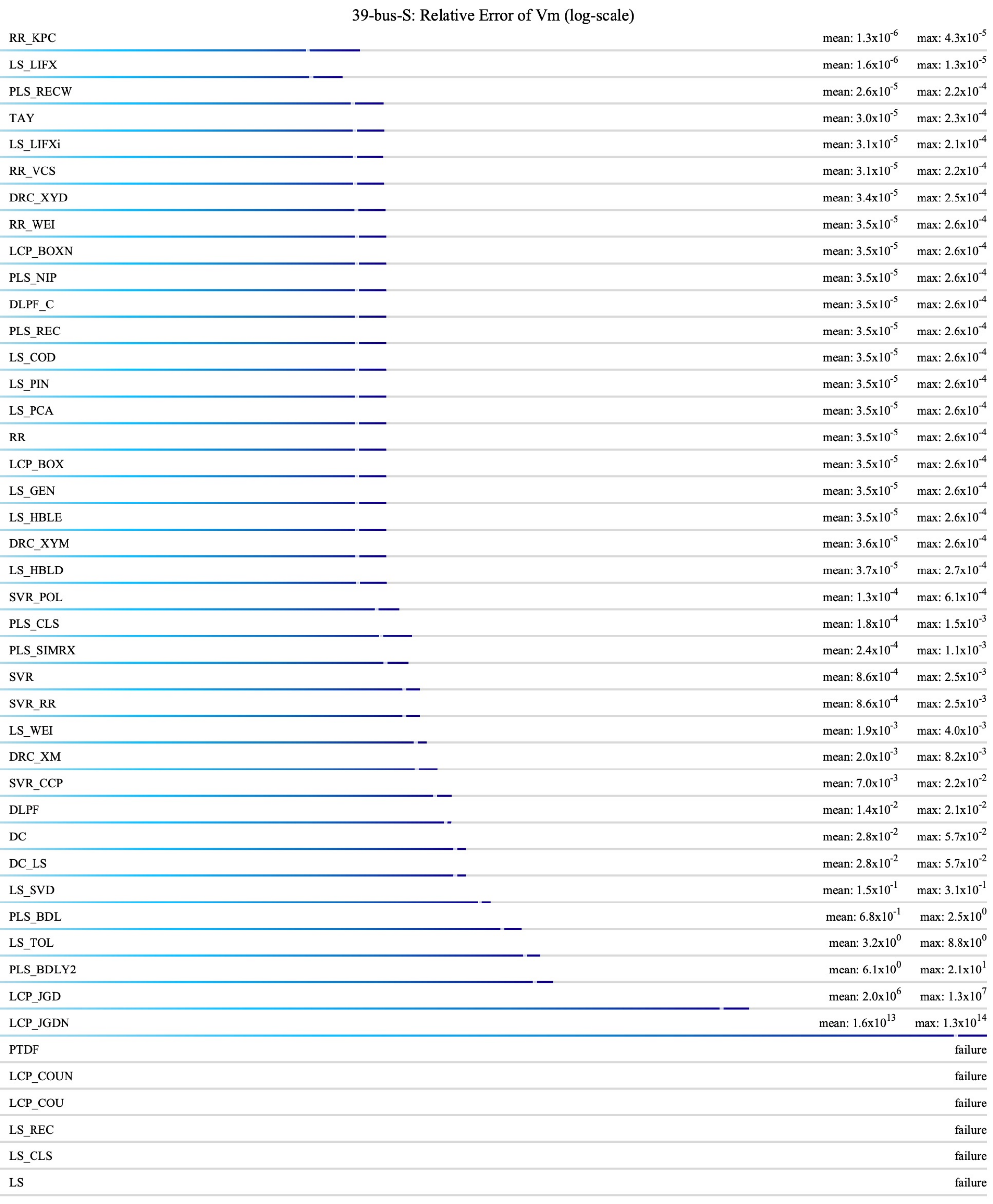} 
	\caption{Linearization accuracy ranking of nodal voltages of all methods, evaluated under the 39-bus-S case. }
	\label{Fig:39-bus-S-Vm} 
\end{figure*}

\begin{figure*}[H]
	\centering 
	\includegraphics[width=7.1in]{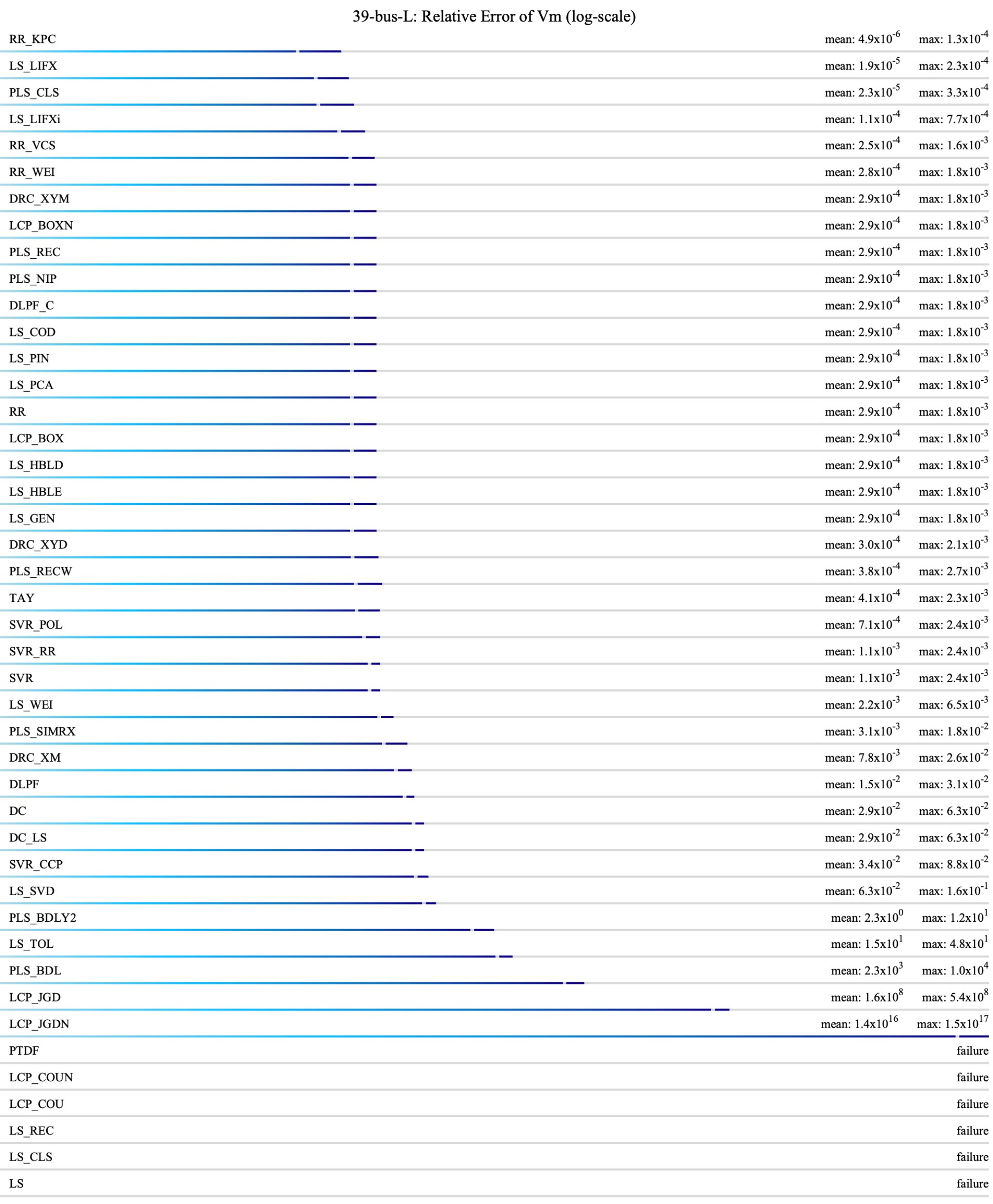} 
	\caption{Linearization accuracy ranking of nodal voltages of all methods, evaluated under the 39-bus-L case. }
	\label{Fig:39-bus-L-Vm} 
\end{figure*}

\begin{figure*}[H]
	\centering 
	\includegraphics[width=7.1in]{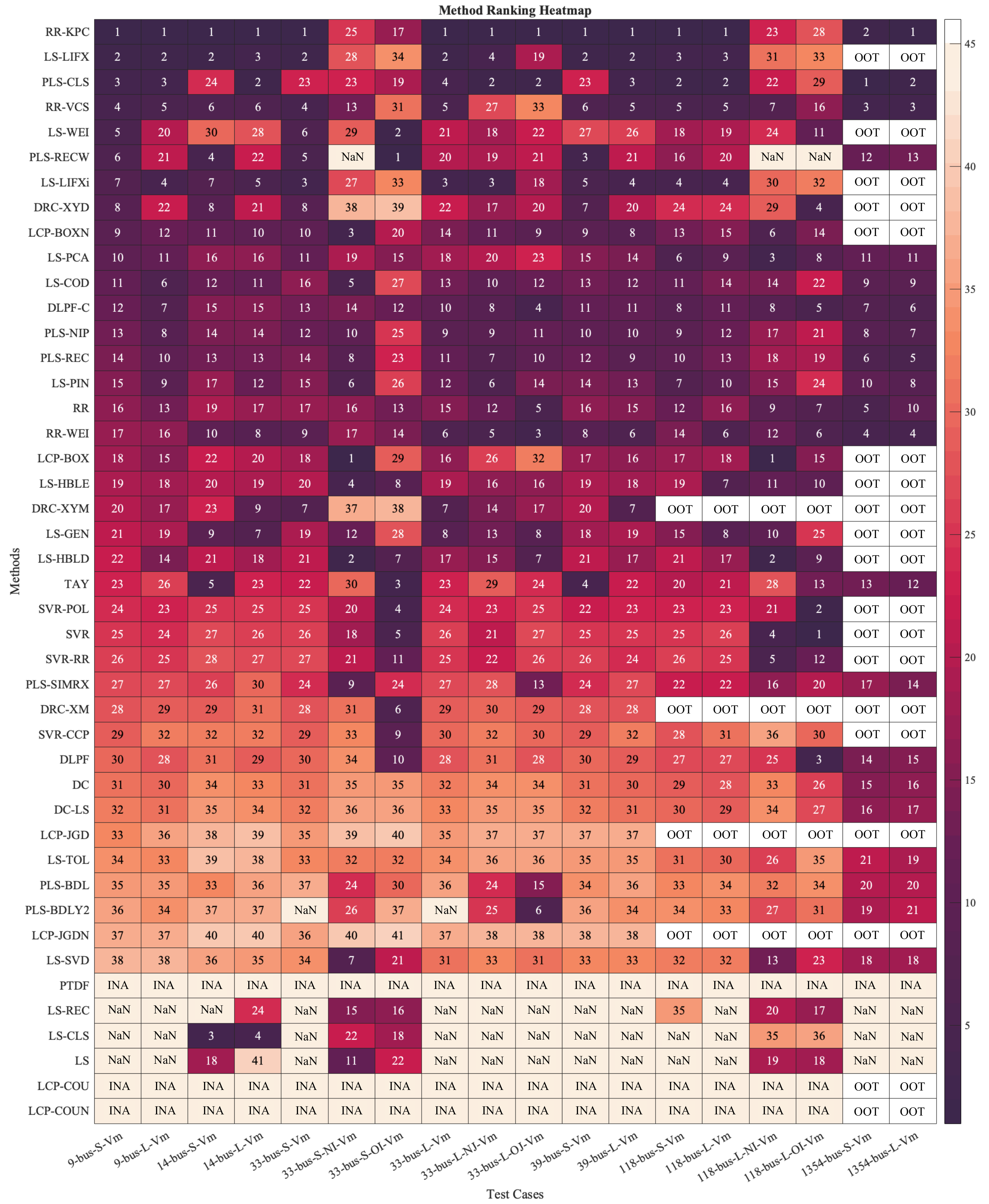} 
	\vspace{-0.4cm}
	\caption{Rankings of the linearization accuracy of 44 distinct methods for voltage magnitudes across various test conditions. The color intensity and labels used here have the same meanings as in Fig. \ref{Fig:rankPF}. See Figs. \ref{Fig:33-bus-S-PF}, \ref{Fig:33-bus-S-NI-PF}, \ref{Fig:39-bus-S-Vm}, \ref{Fig:39-bus-L-Vm} and those in the supplementary file \cite{supplement} for the specific ranking results, including mean and maximum relative errors for every method under each test case.}
	\label{Fig:rankVm} 
\end{figure*}

Detailed error comparisons are presented in Figs. \ref{Fig:33-bus-S-PF}, \ref{Fig:33-bus-S-NI-PF}, \ref{Fig:39-bus-S-Vm}, and \ref{Fig:39-bus-L-Vm}. For instance, in Fig. \ref{Fig:33-bus-S-PF}, the mean relative error of the most precise PPFL method is significantly higher --- by five orders of magnitude --- than that of the leading DPFL approach. Similarly, in the test cases of Fig. \ref{Fig:33-bus-S-NI-PF} and \ref{Fig:39-bus-L-Vm}, the top PPFL method's mean relative error remains two orders of magnitude larger than that of the leading DPFL approach. The gap narrows only in Fig. \ref{Fig:39-bus-S-Vm}, where the mean relative error of the best PPFL method is one order of magnitude greater than that of the best DPFL method.

It is important to highlight that although numerous tests conducted in our study indicate a general trend of DPFL approaches outperforming the accuracy of widely utilized PPDL methods, this paper does not claim that DPFL methods are always superior. The primary takeaway from our extensive testing is that DPFL approaches can exhibit accuracy levels comparable with or even higher than those of PPDL methods, emphasizing their potential relevance and the need for further exploration in this field.

\subsubsection{DPFL Performance: General View}
Addressing the inherent nonlinearity of AC power flows is crucial for improving linearization accuracy. As illustrated in Figs. \ref{Fig:rankPF} and \ref{Fig:rankVm}, clustering-based and coordinate-transformation-based DPFL methods generally exhibit higher accuracy than the rest. The clustering-based methods include RR\_KPC and PLS\_CLS, which produce type-2 (piecewise linear) models, while the transformation-based methods contain LS\_LIFX, LS\_LIFXi, and RR\_VCS, which generate type-3 (linear models only in a transformed space) models. These results show the effectiveness of both clustering and coordinate transformation in handling the nonlinearity of AC power flows, as discussed in \cite{partI}. However, the following points should be noted:
\begin{itemize}
    \item The enhanced precision of the clustering-based and coordinate-transformation-based DPFL approaches is acquired at a cost: the models generated by these methods may not be inherently suitable for practical decision-making applications. E.g., for piecewise linear models, there is a requirement for extra efforts (e.g., the introduction of additional integer variables) to facilitate their integration into optimization frameworks, which increases the complexity of the decision-making framework. Moreover, models derived through coordinate transformation are no longer linear representations of conventional variables but are linear representations of nonlinear mappings of these variables. Consequently, such models might be incompatible with the framework of decision-making problems such as unit commitment. 
    \item Method PLS\_CLS, unsophistically designed in this paper to simply illustrate the modular architecture of DPFL, demonstrates surprising accuracy with a superior ranking. Given its significantly lower complexity compared to the RR\_KPC approach, the performance of PLS\_CLS implies the considerable potential inherent in investigating the optimal ensemble strategy for the architectural configuration of DPFL.
    \item In the presence of individual data noise or outliers within the dataset, the reliability of clustering-based and coordinate-transformation-based DPFL approaches is compromised. In these cases, alternative methods such as the method DLPF\_C for voltage calculation, or the methods PLS\_NIP, PLS\_REC, DLFP\_C, LS\_COD, and LS\_PIN for active branch flow calculation, are more reliable. It is important to note that although these methods are ranked between 4th and 15th, the differences in error among their resulting models are minimal. For instance, in the case of 33-bus-S-PF, the mean and maximal relative errors for the methods PLS\_NIP, PLS\_REC, DLFP\_C, LS\_COD, and LS\_PIN are indistinguishable, as shown in Fig. \ref{Fig:33-bus-S-PF}. 
    \item Regardless of the consistency in the rankings of certain methods, there is a notable degradation in the accuracy of all DPFL methods when exposed to individual data noise/outliers. E.g., as observed in Fig. \ref{Fig:33-bus-S-PF} and \ref{Fig:33-bus-S-NI-PF}, the mean and maximal errors for all DPFL approaches increase by one to three orders of magnitude in the case with data noise, as opposed to the noise-free condition. Such error amplification is uniformly observed across all test cases exposed to individual data noise/outliers. These results highlight the importance of the data cleaning process, e.g., filtering noise and outliers prior to data training. 
    \item The results presented in Figs. \ref{Fig:rankPF} and \ref{Fig:rankVm} indicate that individual data noise/outliers yield a more significant influence on the performance of most approaches compared to joint data noise/outliers. This observation emphasizes the importance of transparency in research publications regarding the methodologies employed for introducing noise and outliers into datasets, not only to ensure clarity but also reproducibility. Henceforth, the term ``noise/outliers'' will be used to specifically denote individual data noise/outliers unless otherwise specified.
\end{itemize}

\subsubsection{DPFL Performance: Individual View}

The comparison of all methods reveals extensive information, particularly when focusing on some specific methods. Therefore, in the following, we present performance analyses of particular methods from their individual perspectives.

\paragraph{LS, LS\_CLS, and LS\_REC}~\\

First, the core of these three approaches is the original least squares method, which inherently has difficulties with multicollinearity, leading to frequent failures in our experiments. However, this does not imply a uniform failure across all scenarios, as in some instances, multicollinearity is less apparent. For example, in certain cases like 14-bus-S and 14-bus-L, the original dataset inherently does not have multicollinearity issues. Moreover, the introduction of data noise or outliers on an individual basis for every predictor can mitigate the interdependence among predictors, effectively addressing the multicollinearity issue. In such situations, as observed in 33-bus-S-NI, 33-bus-S-OI, 118-bus-L-NI, and 118-bus-L-OI, the methods LS, LS\_CLS, and LS\_REC did not fail, as indicated in Figs. \ref{Fig:rankPF} and \ref{Fig:rankVm}.
    
Second, the LS\_REC method begins its updates from a model initially derived through the LS approach using a subset of the training data, referred to as the old dataset. In contrast, the LS and LS\_CLS methods utilize the entire training dataset, encompassing both the old and new datasets, for their training. This difference contributes to the inconsistent performance outcomes among these three methods. For instance, in the 14-bus-S test case, LS\_REC encountered failure (as indicated in Figs. \ref{Fig:rankPF} and \ref{Fig:rankVm}), due to multicollinearity within the old dataset. On the other hand, LS and LS\_CLS did not fail as expanding the old dataset with new data eliminates the multicollinearity issue, illustrating that an increase in dataset size can potentially mitigate such issues. However, this phenomenon does not universally apply. For example, in the 118-bus-S case, the old dataset did not exhibit multicollinearity, allowing LS\_REC to perform successfully. Conversely, incorporating new data into the dataset led to LS and LS\_CLS failing due to the emergence of multicollinearity. This observation suggests that while altering the dataset size can influence multicollinearity, it offers no certainty regarding whether it will worsen or lessen the issue. 

\paragraph{LS\_TOL and LS\_SVD}~\\

The performance of the LS\_TOL method falls below average in most cases, primarily because of the singularity issue. As detailed in \cite{partI}, LS\_TOL initiates with singular value decomposition (SVD) and subsequently estimates the linear model by inverting a submatrix $\boldsymbol{V}_{yy}$ extracted from the matrix \(\boldsymbol{V}\), which is a result of SVD. While \(\boldsymbol{V}\) is inherently non-singular, this attribute does not extend to its derived submatrices, such as $\boldsymbol{V}_{yy}$. Frequently, the determinant of $\boldsymbol{V}_{yy}$ is extremely close to zero. The inversion of $\boldsymbol{V}_{yy}$, therefore, results in substantial errors in the linear power flow model obtained by LS\_TOL. Certainly, the root cause of this issue can be traced back to multicollinearity being present within the dataset of electrical measurements.

Likewise, method LS\_SVD exhibited bad performance in test cases. This is primarily due to multicollinearity within the predictor's training data, which can result in the Gram matrix having zero singular values. Consequently, a classical outcome of SVD --- the $\boldsymbol{S}$ matrix --- may be extremely close to singularity. This near-singular condition significantly hinders the inversion accuracy of this matrix when implementing LS\_SVD, leading to substantial errors.

\paragraph{LS\_LIFX and LS\_LIFXi}~\\

The LS\_LIFXi method is the original version proposed in \cite{1}. It specifically involves lifting each dimension of the predictor on an individual basis using different lift-dimension functions, as given in equations (11)-(15) of \cite{1}. Note that LS\_LIFXi draws inspiration from the dimension-lifting concept introduced in \cite{korda2018linear}. Yet, according to our understanding, the approach described in \cite{korda2018linear} jointly lifts all dimensions of the predictor using one lift-dimension function, rather than doing so individually with multiple lift-dimension functions. To respect the original concept in \cite{korda2018linear}, we have also employed the LS\_LIFX method in this paper. As seen from the evaluation in Figs. \ref{Fig:rankPF} and \ref{Fig:rankVm}, LS\_LIFX often demonstrates an improvement over LS\_LIFXi, particularly in scenarios without data noise or outliers. This observation suggests that lifting all dimensions of the predictor jointly may yield enhanced results when the data is clean, though the error difference between the two methods is minimal.

\paragraph{LS\_HBLD and LS\_HBLE}~\\

In theory, methods LS\_HBLD and LS\_HBLE are expected to yield identical outcomes since both methods address a convex optimization problem and the transformations they adopt are equivalent. Practically, the results from LS\_HBLD and LS\_HBLE are also regarded as indistinguishable. Although their rankings might not be consistently next to each other, as illustrated in Figs. \ref{Fig:33-bus-S-PF}-\ref{Fig:39-bus-L-Vm}, the difference in their error magnitudes is minimal and can be attributed primarily to numerical inaccuracies, such as rounding errors.
    
Additionally, the two methods are specifically designed to address data outliers, including significant noise. They indeed have demonstrated improvements in accuracy when exposed to polluted data. For instance, in Fig. \ref{Fig:rankVm}, in the test cases 33-bus-S-OI and 118-bus-L-OI, there is a notable enhancement in the rankings of both methods. Nevertheless, this trend is less apparent in Fig. \ref{Fig:rankPF}. This difference may be attributed to the nature of the data involved: active branch flow data exhibit greater variability and sudden changes compared to voltage data. Consequently, applying a threshold to differentiate between normal data points and outliers, the core idea of LS\_HBLD and LS\_HBLE,  will be more effective for voltage data (Fig. \ref{Fig:rankVm}) than for active branch flow data (Fig. \ref{Fig:rankPF}).

\paragraph{SVR, SVR\_RR, and SVR\_POL}~\\

First, as depicted in Figs. \ref{Fig:rankPF} and \ref{Fig:rankVm}, the performance of methods SVR and SVR\_RR is notably similar in most instances. This similarity primarily stems from the regularization factor for SVR\_RR, which is configured to be as small as that in conventional methods like RR. A larger regularization factor can result in differences. 
    
Additionally, SVR\_POL offers no distinct advantages over SVR for two major reasons. Firstly, in situations where the underlying power flow data structure is nearly linear, the benefit of a polynomial kernel declines, leading to similar performance between SVR\_POL and SVR with a linear kernel. Secondly, kernel selection is inherently challenging; while a 3rd-order polynomial kernel is a standard choice within the reproducing Hilbert kernel space, it may not effectively enable a highly linear representation of the projected power flow model. 
    
Furthermore, methods SVR, SVR\_RR, and SVR\_POL should be robust to data outliers. This holds for the calculation of the voltage, as shown in Fig. \ref{Fig:rankVm}, where these three methods show a large improvement in the rankings under the cases with data outliers, e.g., 33-bus-S-OI and 118-bus-L-OI. However, such an improvement is less obvious when computing active branch flows, as shown in Fig. \ref{Fig:rankPF}. Similar to methods LS\_HBLD and LS\_HBLE, methods SVR, SVR\_RR, and SVR\_POL use a threshold to distinguish between data outliers and normal data points. Such a manner may be ineffective when large fluctuations exist in the dataset, such as the dataset of active branch flows. 

Moreover, methods SVR, SVR\_RR, and SVR\_POL are designed to be robust against data outliers. Their effectiveness is evident in voltage calculations, as shown in Fig. \ref{Fig:rankVm}: in test cases with data outliers, such as 33-bus-S-OI and 118-bus-L-OI, these three methods significantly outperform others. However, this enhancement is less pronounced in the computation of active branch flows, as illustrated in Fig. \ref{Fig:rankPF}. Similar to LS\_HBLD and LS\_HBLE, SVR, SVR\_RR, and SVR\_POL employ a threshold to differentiate between outliers and typical data points. One threshold may be insufficient for datasets with extensive fluctuations, like those encountered in active branch flow data. In other words, effectively tuning this threshold can be challenging for such datasets.

\paragraph{PLS\_BDL and PLS\_BDLY2}~\\

First, as previously explained, the multicollinearity problem is more evident in PLS\_BDLY2 than in PLS\_BDL,  due to the improper placement of the active power injection at the slack bus. This explains the failure of PLS\_BDLY2 in specific situations where PLS\_BDL remains effective.

Second, PLS\_BDL and PLS\_BDLY2 exhibit poor performance as indicated in both Figs. \ref{Fig:rankPF} and \ref{Fig:rankVm}, which can be attributed to the presence of constant columns in the training dataset --- as indicated in Table \ref{tab:cases}, generator terminal voltages are fixed, resulting in these constant columns. Due to their reliance on the bundle strategy, as outlined in \cite{partI}, PLS\_BDL and PLS\_BDLY2 cannot handle these constant features effectively. This is because, again, the presence of constant columns disrupts the invertibility of a crucial matrix within their computational framework, thereby compromising the performance of these methods. However, introducing noise or outliers into the data --- whether individual noise/outliers or joint noise/outliers --- alters this condition, since the constant columns are disrupted by such data irregularities, which improves the invertibility of the aforementioned critical matrix. Consequently, in test cases such as 33-bus-S-NI, 33-bus-S-OI, 33-bus-L-NJ, 33-bus-L-OJ, 118-bus-L-NI, and 118-bus-L-OI, both PLS\_BDL and PLS\_BDLY2 exhibit enhanced performance compared to their results in other cases.

\paragraph{LCP\_COU and LCP\_COUN}~\\

In Fig. \ref{Fig:rankPF}, the superior performance of LCP\_COU over LCP\_COUN can be attributed to the integration of real physical relationships into the LCP\_COU. Specifically, LCP\_COU incorporates the ground truth that the signs of the coefficients corresponding to the terminal angles of a line should be opposite. This physical insight aids in finding a more accurate solution, highlighting the value of embedding real physics into the training for DPFL approaches.

\paragraph{LCP\_BOX and LCP\_BOXN}~\\

Contrary to the above scenario, LCP\_BOXN outperforms LCP\_BOX according to Fig. \ref{Fig:rankVm}. This difference in performance can likely be traced back to the nature of the physical knowledge incorporated into LCP\_BOX. The information embedded in LCP\_BOX is derived from a physics-based linear approximation, which does not equate to actual ground truth physical knowledge. Introducing this approximated physical knowledge does not necessarily enhance accuracy and, as observed in our study, can even be harmful. This suggests that the effectiveness of incorporating physical knowledge into DPFL training is highly dependent on the precision of that knowledge.

\paragraph{LCP\_JGD and LCP\_JGDN}~\\

Both LCP\_JGD and LCP\_JGDN exhibit limited performance in Figs. \ref{Fig:rankPF} and \ref{Fig:rankVm}, primarily due to their adoption of the bundle strategy, akin to PLS\_BDL and PLS\_BDLY2. This strategy compromises their ability to effectively process datasets with constant columns, resulting from fixed voltage values within the data. Nonetheless, the performance of LCP\_JGD and LCP\_JGDN could potentially be improved by omitting the bundle strategy. For example, in the case of 9-bus-S, as shown in the supplementary material \cite{supplement}, applying the bundle strategy leads to a critically low determinant (below $2.3 \times 10^{-38}$) for the matrix requiring inversion. This result in maximum relative errors for LCP\_JGD and LCP\_JGDN of $2.6 \times 10^6$ and $1.9 \times 10^{13}$, respectively. In contrast, the removal of the bundle strategy results in significantly reduced maximum relative errors, standing at $2.6 \times 10^{-1}$ for LCP\_JGD and $4.1 \times 10^{-1}$ for LCP\_JGDN, respectively. This comparison implies the detrimental impact the bundle strategy could have under certain circumstances.

\subsection{Efficiency Evaluation}

The evaluation of computational efficiency in this section includes a comparison of DPFL and PPFL methods, a general review of DPFL methods, and individual examinations of specific groups of DPFL approaches. 


\subsubsection{DPFL Performance vs. PPFL Performance}
\begin{figure*}[H]
	\centering 
	\includegraphics[width=7.1in]{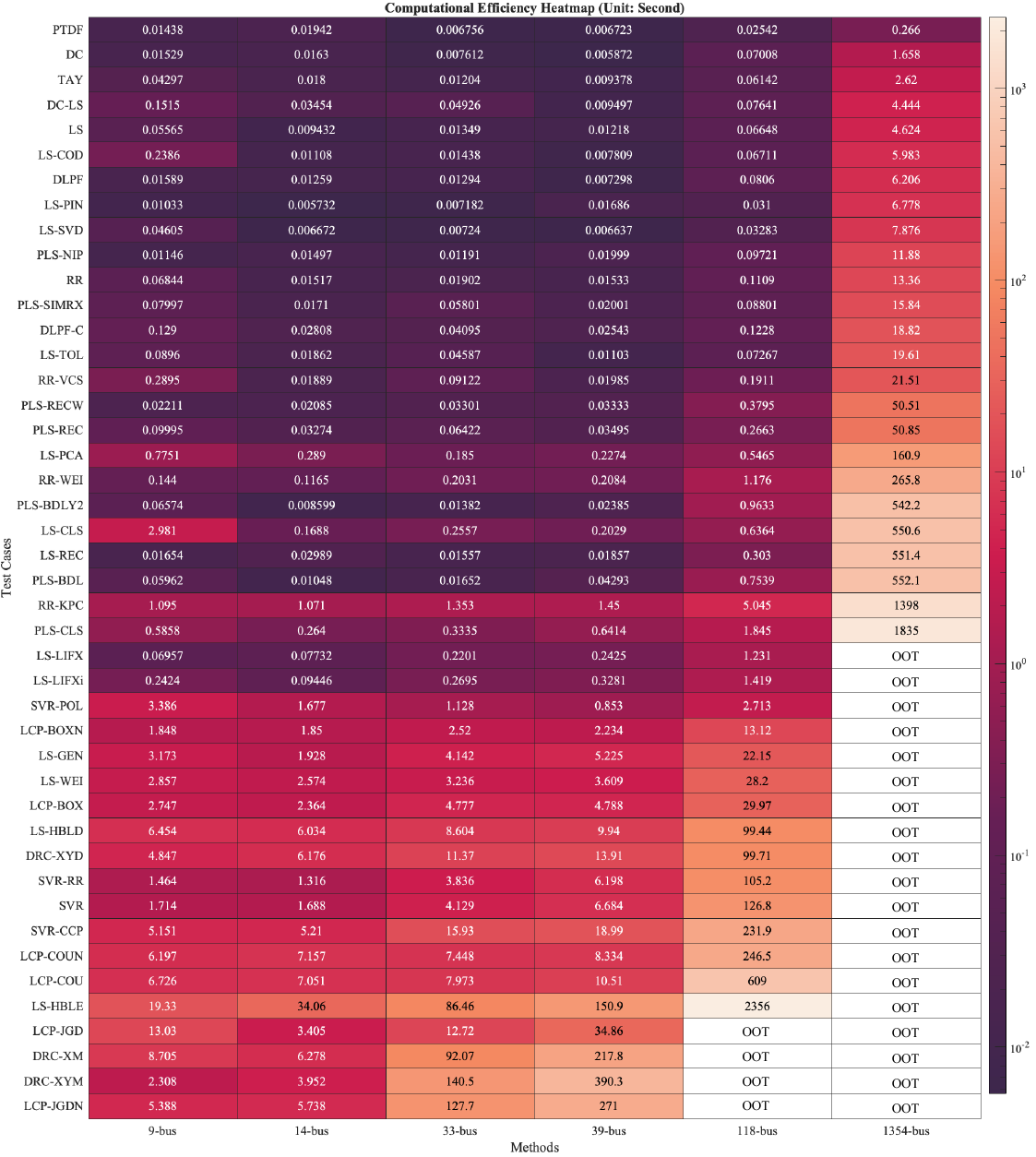} 
	\vspace{-0.4cm}
	\caption{Computational times of evaluated methods under different test cases. These test cases are: 9-bus-S, 14-bus-S, 33-bus-S, 39-bus-S, 118-bus-S, and 1354-bus-S. Note that some methods encountered the OOT-type failure in certain scenarios, as explained before. This experiment employed a consumer-grade laptop equipped with an M1 chip and 16 GB of RAM.}
	\label{Fig:Efficiency} 
\end{figure*}

As illustrated in Fig. \ref{Fig:Efficiency}, the computational efficiency of PPFL methods notably outperforms that of DPFL approaches. This advantage largely stems from the inherent nature of PPFL approaches, i.e., avoiding the need for a training process, in contrast to DPFL methods. Fundamentally, the absence of training in PPFL methods is attributed to their reliance on predefined physical models of the system, assuming that all necessary physical parameters are known and accurate. However, this assumption might not always be valid, particularly in distribution grids. 

\subsubsection{DPFL Performance: General View}

The computational costs of the majority of DPFL methods exhibit superlinear growth with increasing system size, approaching or even exceeding quadratic growth rates, even for methods with closed-form solutions. Much of the computational effort is devoted to model fitting, particularly for DPFL methods that require tuning of hyperparameters. For instance, clustering-based methods have to tune the number of clusters by cross-validation, while ridge-regression-related methods need cross-validation to adjust the regularization factor. This tuning process can be computationally intensive, as it involves systematically searching for better hyperparameter values, which often require multiple iterations of model training and validation. 

It is important to highlight that, aside from the DPFL methods which encountered OOT failures, several DPFL approaches, particularly in the 1354-bus-S case, showed computational times exceeding 300 seconds. Considering that the real-time dispatch in some countries operates on a five-minute cycle, these slower DPFL methods fall short of meeting the real-world computational speed requirements. Given the considerable computational burden many DPFL methods face in large systems, exploring ways to accelerate their training becomes a practical concern. However, it is crucial to clarify that not all DPFL methods suffer from excessive computational efforts. Recursive DPFL methods, in particular, demonstrate more promising speed, which will be further explained in the subsequent discussion.

\subsubsection{DPFL Performance: Individual View}
In the following, we present performance analyses of particular groups of methods.

\paragraph{High-burden Methods}~\\

High-burden methods refer to the methods that result in OOT failures in large-scale test cases. Such a high burden is primarily due to the reliance on optimization programming techniques. Examples include LS\_HBLD, LS\_HBLE, LS\_WEI, SVR, SVR\_CCP, SVR\_RR, DRC\_XYD, DRC\_XYM, DRC\_XM, LCP\_BOX, LCP\_BOXN, LCP\_COU, LCP\_COUN, LCP\_JGD, and LCP\_JGDN. Even for unconstrained quadratic programming, e.g., LCP\_JGDN, the computational demands can quickly become overwhelming and surpass memory capabilities. For a situation where the optimization models have more complex constraints like distributionally robust or chance constraints, the computational demand can be even higher. 
    
On the other hand, LS\_LIFX and LS\_LIFXi, which can also be considered as, fail in large-scale tests as they significantly increase the dataset size by dimension lifting, exceeding MATLAB's maximum array size limitations. 

Additionally, SVR\_POL and LS\_GEN, the last two high-burden methods, experience significant slowdowns at a system size of 1354 buses, a consequence of their specialized design of 3rd-order polynomial-kernel fitting and iterative training, respectively, as explained before.

\paragraph{Clustering-based Methods}~\\

Clustering-based methods like RR\_KPC and PSL\_CLS often outperform other DPFL approaches in accuracy but are notably more time-intensive. For instance, processing a 1354-bus-S case requires over 1000 seconds with these methods. The primary burden lies in the clustering phase, which is particularly demanding for large-scale datasets. This is because both RR\_KPC's K-plane clustering and PSL\_CLS's K-means clustering involve iterative searches for better numbers of clusters, an inherently resource-intensive process. Furthermore, these approaches require the tuning of hyperparameters through cross-validation, adding more layers of iterative computation. E.g., RR\_KPC needs adjustments in both the number of clusters and regularization parameters; such a combinational search plus the clustering process significantly increases the computational demand.

\paragraph{Recursive Methods}~\\

Methods PLS\_REC and LS\_REC are distinguished by their recursive nature, setting them apart from techniques that rely on the entire training dataset for model fitting. Instead,  PLS\_REC and LS\_REC incrementally refine the model through hundreds or thousands of updates. For instance, in the 1343-bus-S case, PLS\_REC begins by using 40\% of the training dataset, which equates to the old dataset with 1200 data samples, to establish an initial linear model via the ordinary partial least squares approach. Subsequently, PLS\_REC imitates an online procedure by incorporating each new data point as it arrives, thereby updating the linear model recursively. Ultimately, PLS\_REC performs 1800 updates to derive the final model, which amounts to the entire training set comprising 3000 data points. Note that the entire recursive process is remarkably efficient, i.e., 50.85 seconds in total or approximately 0.028 seconds for each point-wise update for this test case. Considering a measuring time resolution of one point every half second, where five minutes yield 600 data points, these data can be recursively integrated by PLS\_REC in roughly 16.8 seconds using a consumer-grade laptop. This efficiency aligns well with the requirements of five-minute real-time dispatch scenarios.

\paragraph{Well-rounded Methods}~\\

Methods that balance high accuracy, efficiency, and practicality (where ``practical'' means not significantly increasing the complexity of the application) include LS\_COD, DLPF\_C, PLS\_NIP, and PLS\_REC. While the transformation-based RR\_VCS method has advantages in both accuracy and computational time, its model is challenging to apply in optimization and similar applications.

\section{Open Questions}
The discussions and simulations presented in this paper, alongside \cite{partI}, imply a range of open problems and suggest potential areas for future research. Note that \cite{partI} already dived into the limitations of specific methods, revealing potential open problems from an individual method's view. Hence, this section adopts a broader and complementary perspective, not merely summarizing the overarching issues and directions for future research from a general standpoint, but also highlighting the inconsistency between expected capabilities and actual simulation outcomes for several specific approaches. 

\paragraph{Data Pollution}~\\

How to handle data noise and outliers is still an open question. Current approaches lack reliable methods for effectively dealing with outliers. As discussed above, while SVR-related and Huber-loss-related methods easily detect the outliers in the voltage data, they did not perform well in identifying outliers within the data of branch flows. Similarly, existing methods to mitigate noise, such as using LS\_TOL, do not yield satisfactory results. A more robust solution is needed, one that can filter noise and outliers without the impracticalities of traditional methods like Kalman filtering, which demands assumptions about distribution and fluctuation that are not readily applicable in the dynamic environment of power systems. The challenge lies in developing an approach that not only improves upon the accuracy of detecting outliers and noise in branch flow data but is also adaptable and practical for real-world application.

\paragraph{Suboptimality}~\\

All DPFL methods may exhibit suboptimal performance for several reasons. Firstly, the inherent hyperparameters in many DPFL techniques pose a challenge for parameter optimization. Cross-validation is commonly employed for tuning hyperparameters, but this approach is time-intensive and potentially suboptimal due to the limitation of predefined hyperparameter ranges. Furthermore, DPFL methods exhibit evident modular characteristics, allowing for the integration of various modules to address various issues, such as multicollinearity and inherent nonlinearity. In methods without conventional hyperparameters, the selection or combination of modules can also be seen as a hyperparameter in a general sense, which is more complex to tune and prone to suboptimality. Thus, theoretically defining optimality in DPFL method design and identifying optimal hyperparameters (both conventional and modular) without significantly sacrificing computational efficiency, remain critical yet unresolved areas. Note, again, that the PLS\_CLS method, simply designed to highlight DPFL's modular nature, already demonstrates considerable potential. The potential performance of a DPFL method thoughtfully designed by leveraging its modular advantages is likely to be substantial.

\paragraph{Computation Efficiency}~\\

Over 20 methods are highly time-consuming, especially for large power systems with over 1000 buses. Even though some of these methods are quite accurate, they cannot satisfy the time requirements of several applications such as real-time dispatch. These time-consuming methods are mainly optimization-based, nonlinear-model-fitting-based, or iteration-based approaches. Therefore, enhancing the efficiency of optimization problem-solving, improving the fitting of nonlinear models, accelerating the convergence of specific algorithms, and effectively managing large-scale datasets are critical strategies for addressing the issue of high computational burden. These measures can significantly contribute to the practical applicability and performance of DPFL methods in large scenarios. However, how to achieve such acceleration remains a quite large open area to investigate.

\paragraph{Inherent Nonlinearity}~\\

In addressing the inherent nonlinearity within AC power flow models, coordinate transformation strategies such as LS\_LIFX, LS\_LIFXi, and RR\_VCS have shown promise. However, they compromise the practical applicability of DPFL models for optimization or control purposes, as they render the model a linear function of transformed variables that lack clear physical interpretation. Furthermore, the application of nonlinear kernels to project the AC model into another space, despite its theoretical appeal, presents challenges in identifying an effective nonlinear kernel that can facilitate a highly linear representation of the AC power flow model in the transformed space. The commonly used SVR\_POL with a 3rd-polynomial kernel falls short of achieving such an effective power flow representation. Moreover, piecewise linear models, developed through clustering-based DPFL approaches such as RR\_KPC and PLS\_CLS, demonstrate improved accuracy in cases where the original power flow model exhibits significant nonlinearity. Yet, again,  they are challenging to use in practical applications because they introduce integer decision variables, making the programming problem difficult to solve. Consequently, to handle the inherent nonlinearity of AC power flow models, the investigation to find an optimal balance between model accuracy and practical applicability remains a significant challenge and needs further exploration.

\paragraph{Physical Knowledge}~\\

Several DPFL approaches incorporate various forms of physical knowledge, such as boundary operating conditions (LCP\_JGD), network topologies (LCP\_COU and RR\_VCS), line admittances (LCP\_BOX and LCP\_JGD), and PPFL model formulations (DLPF\_C and DC\_LS). Yet, accessing this physical data may not always be straightforward in real-world scenarios. Future studies need to address this challenge, evaluating the accessibility of such information and reaching a consensus on its availability. Should certain physical data prove difficult to be obtained, assessing the associated costs becomes crucial. This evaluation will help determine the practicality and value of incorporating physical knowledge into DPFL models to enhance their accuracy, considering the potential implications for system operators.

Additionally, integrating physical knowledge into DPFL poses several open questions due to its inherent limitations. First, the process might inadvertently omit critical data by narrowing the focus to certain variables, such as excluding known voltages from the predictors, which could lead to a significant loss of information. Second, the approach faces challenges with normalized datasets, especially when variables are not scaled uniformly, potentially hindering the training process. Third, the effectiveness of this integration heavily depends on the accuracy of the physical knowledge itself, which is not always guaranteed to be precise or universally applicable, as seen with the use of the Jacobian matrix, which is just an approximation of reality. This introduces doubts about the overall enhancement of DPFL model accuracy through physical knowledge, as shown in the simulations. Fourth, the constraints on selecting predictors and responses due to the integration of physical knowledge limit the DPFL model's application scope. The above limitations highlight the need for a critical examination of the role and implementation of physical knowledge in DPFL models, questioning the balance between enrichment and potential data exclusion, and the need for more inclusive and accurate modeling approaches.

\paragraph{Remedial Action} ~\\

In practical scenarios, operating systems often require various remedial measures like topology modifications or adjustments in phase-shifting angles. Models trained from the data prior to these adjustments may lose relevance post-implementation. Thus, incorporating remedial actions as predictors rather than parameters during the model training phase could enhance applicability. However, how to integrate such actions as predictors remains an unresolved question. Furthermore, updating models in light of frequent remedial interventions presents an additional challenge, particularly when transitional data is sparse. With the expected increase in the frequency of remedial actions due to greater integration of renewable energy sources, there is a pressing need for developing a more robust solution. This solution should be grounded in solid theoretical principles to effectively incorporate remedial actions into DPFL models.

\vspace{2cm}
\paragraph{Bus-type Variation} ~\\

Bus-type variation is a notable issue in power flow models. Yet, in DPFL, the current solution to this issue, particularly the bundle strategy employed in methods like PLS\_BDL, PLS\_BDLY2, and PLS\_JGD, faces significant challenges. One major issue is the potential non-invertibility of a key matrix within this strategy's computational framework. Situations where this matrix becomes non-invertible, often due to zero or constant predictors, severely limit the methods' applicability and effectiveness, as evidenced in simulation results. How to effectively address bus-type variations remains a critical area for improvement, especially the need for a more robust solution to ensure the reliability and utility of DPFL approaches.

\paragraph{Limited Observability} ~\\

In scenarios where the system is not fully observed, current DPFL methods are constrained to modeling only those variables that are measured, leading to models that are truncated and limited in their comprehensiveness. The challenge of developing a DPFL model that covers the entire system with only partial observations remains unsolved. Identifying a strategy that overcomes this limitation is crucial, especially for distribution grids where the data availability is often limited.


\paragraph{Test Standardization}~\\

The field of DPFL is currently hindered by the absence of standard testbeds grounded in real-world measurements. To conduct a reliable analysis of DPFL techniques, it is essential to establish a standardized system based on the data from actual power grids. Such a resource would be invaluable for both researchers and practitioners in the DPFL domain, allowing for the evaluation and comparison of various methods in a consistent framework that mirrors real-world conditions. Such a dataset can advance DPFL research and ensure its applicability and effectiveness in practical settings. 
  
\paragraph{DPFL Toolbox}~\\

Beyond the need for standard testsets, the ability to benchmark against established DPFL methods is also vital for researchers to validate their findings. Moreover, providing straightforward access to established DPFL methods also empowers researchers to effortlessly obtain accurate linear models for their projects. Classic PPFL methods such as DC or PTDF are already built-in in many toolboxes, e.g., MATPOWER. However, replicating existing DPFL benchmarks is still a significant challenge due to the lack of open-source codes — over 95\% of the DPFL literature does not provide accessible codes. This gap emphasizes the need for a comprehensive, open-source DPFL toolbox. Such a toolbox not only facilitates the application of all DPFL methods but also enables easy customization of tunable hyperparameters. It should offer a suite of features including data generation, processing, method evaluation, and diverse visualization options. Despite its broad functionality, the toolbox must prioritize user-friendliness; ideally, complex tasks such as method comparison and hyperparameter tuning could be executed with minimal coding effort. The creation of this toolbox would represent a significant leap forward, providing a robust platform for DPFL research and applications, while pushing the boundaries of DPFL into new horizons.

\section{Conclusion}
This paper, as the second part of a tutorial series, presents a comprehensive review of existing DPFL experiments, detailing the capabilities and limitations of these experiments. Additionally, it provides an in-depth numerical examination of DPFL methods, complementing the theoretical insights explored in Part I and addressing the limited experimental work found in the existing literature. This dual focus aims to offer a more holistic understanding of DPFL by bridging the gap between theory and practice.

Specifically, this paper analyzes all the approaches' generalizability towards the selection of predictor and response, as well as their applicability towards multicollinearity, zero predictor, constant predictor, and normalization. Then, through rigorous numerical simulations involving 44 different methods across numerous test cases scaling from 9-bus to 1354-bus, this paper has illustrated the practical performance that mere theoretical analyses could not reveal. The practical performance is evaluated from two angles: accuracy and computational efficiency. For the accuracy evaluation, this paper (i) clarifies the reasons behind the failures encountered by DPFL methods, (ii) provides insights into how DPFL methods compare with traditional PPFL approaches, (iii) offers comprehensive discussions on the uniform outcomes observed across various test cases for all DPFL methods, and (iv) dives into extensive analyses of individual DPFL methods, highlighting their distinctive performances. Regarding the evaluation of computational efficiency, the paper (i) explores the comparative efficiency of DPFL versus PPFL methods, (ii) offers in-depth discussions on the consistent performance metrics of DPFL methods across multiple test scenarios, and (iii) presents detailed evaluations of specific groups of DPFL methods, focusing on their similar performance. Furthermore, drawing on insights from this study and Part I \cite{partI}, as well as the identified open questions, this paper outlines ten promising yet challenging directions, which sketch a roadmap for future research in DPFL. Addressing these questions will not only advance our understanding of DPFL methods but also contribute to the broader goal of developing more reliable, efficient, and accessible tools for power systems research, education, and applications. 

As for our next step, by leveraging our comprehensive repository of documentation and codebase for DPFL methods, we will start with the tenth suggested direction: creating an exhaustive, elegant, and user-friendly toolbox. This toolbox will fully support all DPFL methods and the related functionalities, aiming to offer significant ease of use for researchers and engineers, to support them in exploring and/or leveraging the area of DPFL, thereby reshaping the frontiers of the DPFL research and application.

\section*{Appendix A}

Introducing the principal component analysis into regression can also handle the multicollinearity issue for DPFL. Specifically, as in \cite{partI}, here we again denote the dataset of predictors as $\boldsymbol{X}$, the dataset of responses as $\boldsymbol{Y}$, and the coefficient matrix in the linear model as $\boldsymbol{\beta}$. The principal component analysis can decompose the covariance of $\boldsymbol{X}$, defined as $\rm{cov}(\boldsymbol{X})$ here, into $\rm{cov}(\boldsymbol{X}) = \boldsymbol{D}\boldsymbol{\Lambda}\boldsymbol{D}^{\top}$, where $\boldsymbol{D} \in \mathbb{R}^{N_x \times N_x}$ consists of the eigenvectors of $\rm{cov}(\boldsymbol{X})$. Importantly, by using $\boldsymbol{D}$, the columns in matrix $\boldsymbol{X}\boldsymbol{D}$ not only preserve the information of $\boldsymbol{X}$ but also become uncorrelated. Accordingly, the least squares solution of the principal component regression model, i.e., $$\min \limits_{\boldsymbol{\beta}} \   \Vert \boldsymbol{Y}- \boldsymbol{X}\boldsymbol{D}\boldsymbol{\beta}^{PCA} \Vert_{F}^2 $$
can be given by
$$
\hat{\boldsymbol{\beta}}^{PCA} = \left[ (\boldsymbol{X}\boldsymbol{D})^{\top}(\boldsymbol{X}\boldsymbol{D})\right]^{-1}(\boldsymbol{X}\boldsymbol{D})^{\top}\boldsymbol{Y} \notag
$$
Eventually, the estimation of $\boldsymbol{\beta}$ is found as $\hat{\boldsymbol{\beta}} = \boldsymbol{D}\hat{\boldsymbol{\beta}}^{PCA}$.





\bibliographystyle{elsarticle-num} 
\bibliography{cas-dc-template}




\end{document}